\documentclass[dvipsnames,format=sigconf]{acmart}

\usepackage{mathtools}
\usepackage{mathrsfs}
\usepackage{pifont}

\usepackage{natbib}
\usepackage{url}
\usepackage{booktabs}
\usepackage{hyperref}
\usepackage{cleveref}
\usepackage{stfloats}
\usepackage{subcaption}
\usepackage[space]{grffile}
\usepackage{bm}
\crefname{apsec}{appendix}{appendices}
\Crefname{apsec}{Appendix}{Appendices}

\usepackage{xcolor}
\definecolor{dark-blue}{rgb}{0.05,0.1,0.75}
\hypersetup{
  colorlinks,
  linkcolor={dark-blue},
  citecolor={dark-blue},
  urlcolor={dark-blue}
}

\usepackage{todonotes}
\setuptodonotes{
  inline,
  linecolor=blue,
  backgroundcolor=blue,
  textcolor=white,
  bordercolor=blue,
}

\newcounter{num}

\usepackage{siunitx}
\usepackage{longtable} 

\graphicspath{{./resources}} 

\usepackage[indLines=false]{algpseudocodex}
\usepackage{algorithm}
\makeatletter
\algnewcommand\algorithmiconce{\textbf{Once}}%
\algdef{SE}[ONCE]{Once}{EndOnce}[1]{\algpx@startIndent\algpx@startCodeCommand\algorithmiconce\ #1\ \algorithmicdo%
}{\algpx@endIndent\algpx@startEndBlockCommand\algorithmicend\ \algorithmiconce}%
\pretocmd{\Once}{\algpx@endCodeCommand}{}{}
\algtext*{EndOnce}%
\apptocmd{\EndOnce}{\algpx@endIndent}{}{}%
\pretocmd{\EndOnce}{\algpx@endCodeCommand[1]}{}{}%

\algnewcommand\algorithmicWith{\textbf{With}}%
\algdef{SE}[WITH]{With}{EndWith}[1]{\algpx@startIndent\algpx@startCodeCommand\algorithmicWith\ #1\ \algorithmicdo%
}{\algpx@endIndent\algpx@startEndBlockCommand\algorithmicend\ \algorithmicWith}%
\pretocmd{\With}{\algpx@endCodeCommand}{}{}
\algtext*{EndWith}%
\apptocmd{\EndWith}{\algpx@endIndent}{}{}%
\pretocmd{\EndWith}{\algpx@endCodeCommand[1]}{}{}%
\makeatother

\title{Evolution of Fear and Social Rewards in Prey-Predator Relationship}

\author{Yuji Kanagawa}
\email{yuji.kanagawa@oist.jp}
\orcid{1234-5678-9012}
\author{Kenji Doya}
\authornotemark[1]
\email{doya@oist.jp}
\affiliation{%
  \institution{Okinawa Institute of Science and Technology Graduate University}
  \city{Onna-son}
  \state{Okinawa}
  \country{Japan}
}

\renewcommand\footnotetextcopyrightpermission[1]{} 
\begin{document}

\begin{abstract}
Fear is a critical brain function that enables us to learn to avoid danger via reinforcement learning (RL). While many researchers have argued that fear has evolved to escape predators, how varying predatory pressures have shaped fear and other rewards, including positive social rewards for collective grouping, remains an open question. In this study, we investigate the relationship between predatory pressure and fear using an evolutionary simulation of RL agents with evolving rewards. In our simulation, prey and predator RL agents co-evolve their reward functions, including visual rewards for observing prey and predators. While fear-like negative visual rewards for predators often evolved in prey, we also observed cases in which positive rewards for both predators and prey evolved, the latter serving as a social reward for collective grouping.
A comparison between different environmental conditions revealed that stronger predator hunting capability promoted stronger fear reward, 
while less food supply promoted more negative social reward. Moreover, fear did not evolve in response to static pitfalls with non-lethal damage, suggesting that actively hunting predators played an important role in its evolution. These results highlight the special role of predators in the diverse evolution of fear and social rewards.
\end{abstract}

\maketitle
\section{Introduction}\label{sec:intro}
Fear is a fundamental brain function that helps creatures detect and avoid potential dangers~\citep{adolphsBiologyFear2013}.
Fear as a negative reward enables reinforcement learning (RL) to avoid actions associated with adverse outcomes~\citep{tovoteNeuronalCircuitsFear2015}.
Fear serves as a fundamental part of our brains' reward system~\citep{schultzNeuronalRewardDecision2015}, alongside other rewards such as food and social rewards.

Because animals show fear-related expression to external predators~\citep{apfelbachEffectsPredatorOdors2005,yilmazRapidInnateDefensive2013}, predators are believed to be influential in fear evolution~\citep{marksFearFitnessEvolutionary1994,ohmanFearsPhobiasPreparedness2001}. Because individuals cannot learn from death, an innate fear of predators is necessary for survival. However, because it is difficult to experimentally replicate evolution, it remains unclear how different predatory pressures and their interactions with other rewards affect the evolution of fear.

To this end, we developed a distributed evolutionary simulation framework to reproduce the evolution of fear by simulating prey and predators, building on previous work on food-reward evolution in RL agents with birth and death models~\citep{kanagawaEvolutionRewardsFood2024}. In our simulation, we model fear as a reward for prey's vision-like proximity sensors that detect predators, which can be both negative and positive. Both prey and predators inherit their reward functions from their parents, as encoded in their genotypes, which include components related to food intake, motor output, predator observation, and prey observation.
By replicating birth, death, and reproduction driven by energy metabolism, we expect the reward functions of prey and predators to co-evolve in a biologically plausible manner.

Our simulation results show that fear-like negative rewards for observing predators commonly evolved in prey, although in some cases, positive rewards for both predator and prey evolved, with the latter serving as the social reward for collective grouping.
In comparing different predatory and environmental pressures, we found that enhanced predator hunting capabilities intensified the evolution of fear, while reduced environmental resources had little impact on fear, as they made the environment harsher for both predators and prey. Lastly, we found that fear did not evolve toward static pitfalls with non-lethal damage, suggesting that actively hunting predators can be a primary source for fear evolution. These results indicate that the evolution of fear is closely tied to the interplay between predation and social rewards, suggesting that more complex emotions have also evolved in the social context.

\section{Related Works}\label{sec:related}
Since Darwin observed that Galapagos finches that evolved without predators lacked a fear of humans \citep{darwinVoyageBeagle2022}, the relationship between fear and predation has been a subject of study for centuries. The impact of fear is complex, as it can reduce foraging activity in animals \citep{zanetteEcologyFear2019}. Researchers have demonstrated that fear responses, such as altered birdsong, correlate with population declines \citep{zanetteEcologyNeurobiologyFear2020, allenFearPredatorsFreeliving2022}. Mathematical models grounded in Lotka-Volterra dynamics \citep{wangerskyLotkaVolterraPopulationModels1978} suggest that strong fear responses can actually stabilize populations \citep{sarkarImpactFearEffect2020}. While these mathematical studies typically model fear as a demographic modifier that reduces birth and death rates, our approach employs an agent-based model in which fear serves as a reward for RL agents.

The concept of social reward is often examined in the context of family relationships, including kin selection~\citep{fosterKinSelectionKey2006}. Still, several works argue that predation risks amplify grouping behavior in animal studies~\citep{kramsIncreasedRiskPredation2009} and computer simulations~\citep{olsonExploringEvolutionTradeoff2015}.

We employ a co-evolution model of prey and predator RL agents. While existing studies focus on learning in prey-predator dynamics \citep{olsenCoevolutionPredatorPrey2015,parkCoEvolutionPredatorPreyEcosystems2021} or the evolution of physical properties of agents \citep{yamadaEvolutionComplexPredatorPrey2020}, we focus on the evolution of fear and social rewards.

We built our simulation on previous work that investigated the evolution of food and action rewards~\citep{kanagawaEvolutionRewardsFood2024}. They employed a distributed evolution model inspired by the embodied evolution literature~\citep{bredecheEmbodiedEvolutionCollective2018}, in which each agent maintains an energy level and can die or produce children depending on it. We extend their simulation model to a prey-predator setting.

\section{Simulation Model and Environment}\label{sec:method}

\begin{figure}[t]
\includegraphics[width=8.4cm]{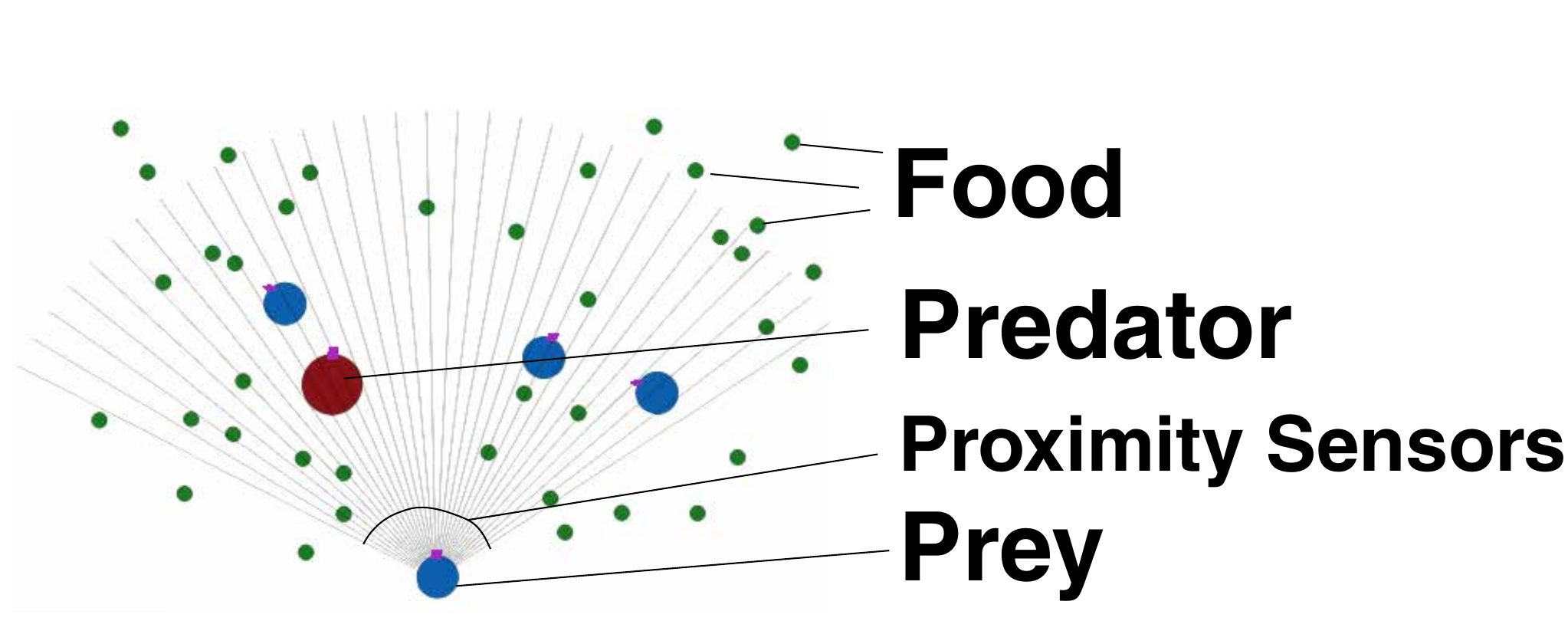}
\caption{
Model of prey and predators used in our simulation. Blue circles represent prey, red circles represent predators, and green circles represent food items that only prey can eat. Gray lines extending from the prey are proximity sensors.
}~\label{figure:zoom}
\end{figure}

\begin{figure}[t]
  \centering
    \begin{subfigure}[b]{0.48\linewidth}
      \centering
      \includegraphics[width=4cm]{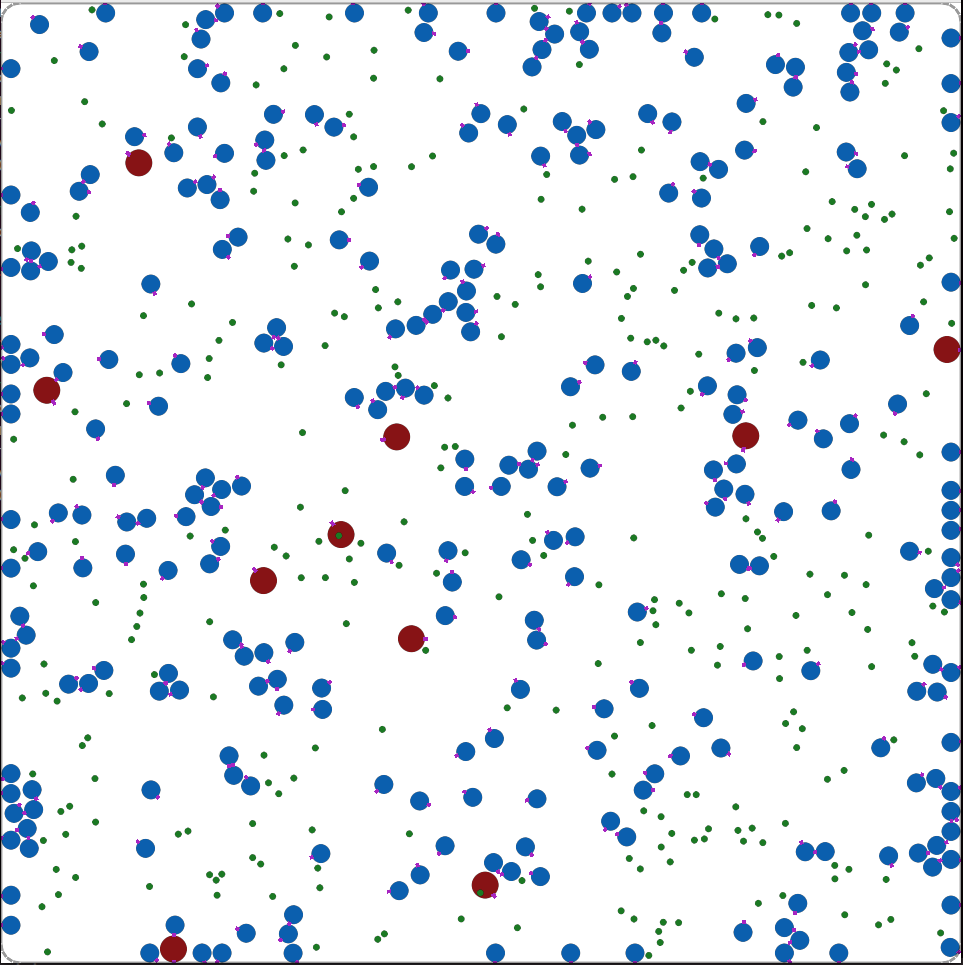}
      \caption{}\label{subfig:ppenv}
    \end{subfigure}
    \hfill
    \begin{subfigure}[b]{0.48\linewidth}
      \centering
      \includegraphics[width=\textwidth]{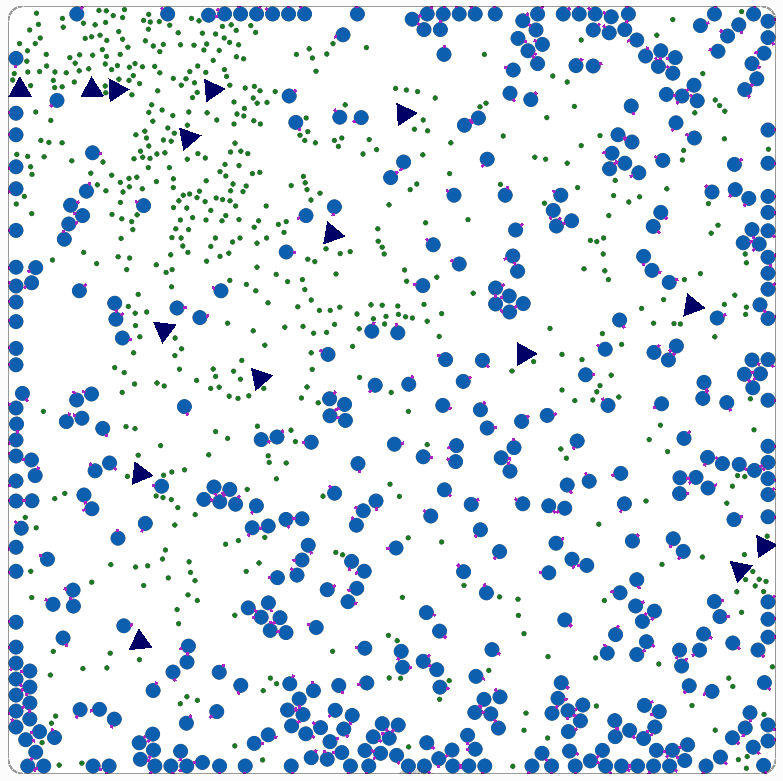}
      \caption{}\label{subfig:obsenv}
    \end{subfigure}
  \caption{(a): Environment with prey and predators.~(b) Environment with prey static pitfalls (black triangles).}~\label{figure:env}
\end{figure}

\begin{figure}[t]
  \centering
  \includegraphics[width=5cm]{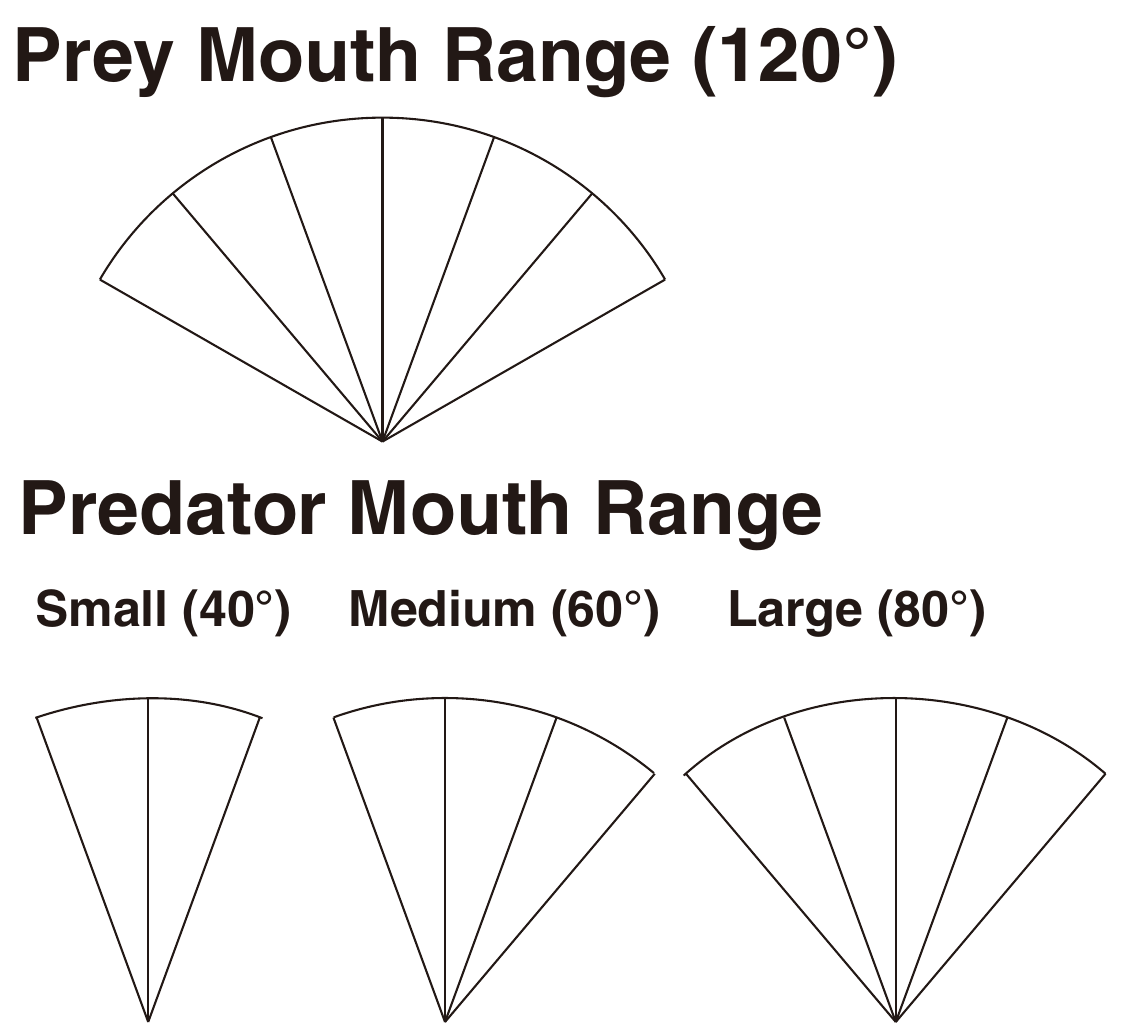}
  \caption{Mouth ranges of prey and predators. M is used as the default value in most experiments.}~\label{figure:mouth}
\end{figure}

Our simulation model extends previous work~\citep{kanagawaEvolutionRewardsFood2024} by introducing two prey and predator species. As illustrated in \Cref{figure:zoom}, both agents are modeled as circular entities within a two-dimensional rigid-body physics environment, with predators being $1.4$ times larger than prey in diameter. Prey can consume green food items by making contact within a $120$-degree forward range, which yields $+1$ energy per item. Predators, conversely, hunt prey by initiating contact within a specific range ($40$ to $80$, as shown in \Cref{figure:mouth}) but remain unable to interact with or consume green food items. Food items are regenerated at random locations up to a fixed capacity of $600$, governed by a regeneration rate $\Delta n$. For most experiments, we set $\Delta n = 0.5$, resulting in the regeneration of one food item every two simulation steps. The agents co-exist within an environment bounded by square walls as shown in \Cref{subfig:ppenv}, with populations capped at $450$ prey and $50$ predators.

Both prey and predators are equipped with $32$ proximity sensors distributed across a $120$-degree forward arc. These sensors measure the inverse distance to the nearest object, scaled to $[0, 1]$. Specifically, the $k$-th sensor input $s^k$ equals $1$ upon contact and $0$ when no object is detected within range. Additionally, $18$ tactile sensors are positioned at $20$-degree intervals around the body to detect localized contact. Agents can move by applying the force $\mathbf{f} = (f_{left}, f_{right})^\top$ to two rear points. The mapping from sensor inputs to the force vector $\mathbf{f}$ is learned via RL.

To isolate the impact of predatory hunting behavior on the evolution of fear, we designed a control environment containing harmful pitfalls in the absence of predators, as shown in \Cref{subfig:obsenv}. This environment supports up to $500$ agents with the same physical specifications as the prey in the prey-predator simulation. The pitfalls, visualized as black triangles, remain stationary across random seeds. Agents lose $10$ energy units upon collision with a pitfall, which can be critical for them.

For each agent $i$, energy level $e^i$ is updated at each time step $j$ by:
\begin{align}
  \Delta{e}_{j}^i =
  \begin{cases}
    n_{j}^{i} - c_{a}\|\bm{f}_{j}^{i}\| - c_{b} & \text{if } i \in \text{Prey} \\
    \sum_{k \in \text{Prey eaten by }i} \eta e^k - d_{a}\|\bm{f}_{j}^{i}\| - d_{b} & \text{if } i \in \text{Predator}
  \end{cases},
  \label{eq:energy}
\end{align}
where $n_{j}^{i}$ is the number of food items acquired and $\|\bm{f}_{j}^{i}\|$ is Euclidean norm of the motor output vector. $c_a$ and $d_a$ are the energy consumption coefficient by motor action, and $c_b$ and $d_b$ are the basal metabolism for prey and predator. $\eta$ is the predator's digetstive rate. Since we use $0.6$ and prey is expected to maintain $10$ to $20$ energy units to survive, predators gain $6$ to $10$ energy units per predation event.

We employ a stochastic birth-and-death model based on metabolic processes to enable distributed evolution under realistic survival pressures. Agents die if their energy level $e$ falls below zero or according to a probability $h(t, e)$, defined by the hazard function:
\begin{align}
  h(t, e) = \kappa_{h}\left(1 - \frac{1}{1 + \alpha_{e}\exp(-\beta_{h}e)}\right)\alpha_{t} \exp(\beta_{t} t),
  \label{eq:h}
\end{align}
where $t$ represents the age of the agent. The age-dependent term $\alpha_{t} \exp(\beta_{t} t)$ accounts for natural longevity.

Both prey and predators reproduce asexually based on a birth probability $b(e)$, which is a function of their current energy level:
\begin{align}
 b(e) &= \frac{\kappa_{b}}{1 + \exp(\zeta - \beta_{b}e)}.
 \label{eq:b}
\end{align}
We set the parameters so that prey require between $20$ and $30$ units of energy for reproduction, whereas predators require a much higher range of $240$ to $260$ units. All parameters used across these experiments are detailed in \Cref{appendix:param}.

While many sensory signals~\citep{kangCentralAlarmSystem2022}, including olfactory~\citep{apfelbachEffectsPredatorOdors2005} signals, are used to elicit innate fear, we chose to use a vision-like proximity sensor as a source of fear because of its ease of implementation. Thus, we model the rewards the rewards $r^{i}_{j} $ for an agent $i$ at step $j$ as follows:
\begin{align*}
  r^{i}_{j} = &w_{\mathrm{eat}}^{i}n_{j}^{i} + 0.01 w_{\mathrm{act}}^{i}\frac{1}{F} \|\bm{f}_{j}^{i}\| \\
  &+ 0.1 w_{\mathrm{prey}}^{i} \max_k{s}_{\mathrm{prey}}^{i, k} + 0.1 w_{\mathrm{pred}}^{i} \max_k{s}_{\mathrm{pred}}^{i, k},
\end{align*}
where $n_{j}^{i}$ represents the number of food items or prey consumed at this step, $|a_{j}^{i}|$ is the magnitude of the normalized motor output, and $F$ is the maximum value of $\|\bm{f}\|$. $\max_k{s}_{\mathrm{pred}}^{i, k}$ is the maximum proximity sensor input, which reflects the closeness to the closest predator. $\max_k{s}_{\mathrm{prey}}^{i, k}$ is the same for prey agents. The reward weights $w_{\mathrm{eat}}$, $w_{\mathrm{act}}$, $w_{\mathrm{prey}}$, and $w_{\mathrm{pred}}$ are genetic parameters that gradually evolve by inheriting values from parents with some mutation. Because $n_{j}^{i}$ is often zero, other reward weights are scaled to take smaller values. The mutation is performed by adding noise sampled from a Student's t-distribution with 2 degrees of freedom and a scale of $0.4$. Because this distribution is heavy-tailed to enhance genetic value exploration, all reward weights are clipped into $[-100, 100]$ after mutation for numerical stability, while we found this clipping doesn't limit the evolution in most cases. In the experiment with stationary pitfalls, we use reward weights $w_{\mathrm{agent}}$ for observing other individuals and $w_{\mathrm{pit}}$ for observing pitfalls, instead of $w_{\mathrm{prey}}$ and $w_{\mathrm{pred}}$ used in the predator-prey environment.

\begin{figure}[t]
\includegraphics[width=8.4cm]{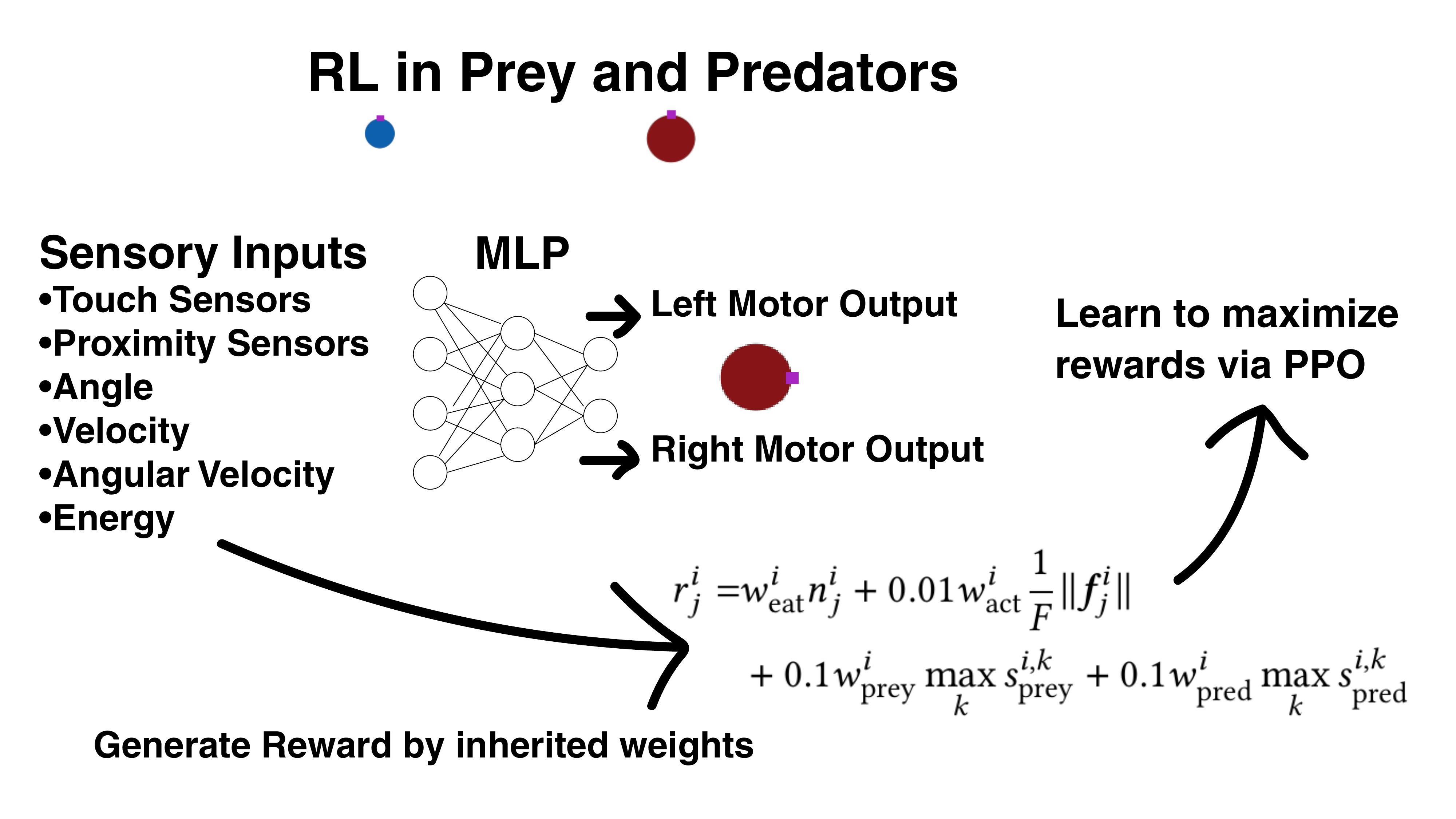}
\caption{Schematic figure shows the learning procedure of prey and predators.}~\label{figure:nn-rl}
\end{figure}

Each agent has its own independent three-layer multi-layer perceptron (MLP) with $64$ hidden units. As shown in \Cref{figure:nn-rl}, the MLP takes as input touch sensors, proximity sensors, angle, velocity, angular velocity, and energy, and outputs policies for two motors as a Gaussian distribution with state-independent variance. Agents act by sampling motor output from this Gaussian policy. A reward is generated from the genetically inherited reward weights, combined with information about food intake, actions, and sensor inputs. From this reward and the history of actions and observations, they learn actions to maximize the sum of rewards via proximal policy optimization (PPO, \citet{schulmanProximalPolicyOptimization2017}). The learned policy is not inherited by newborns. Instead, they all need to learn their behavior from scratch with randomly initialized neural networks.

\begin{figure}[t]
  \centering
  \includegraphics[width=\linewidth]{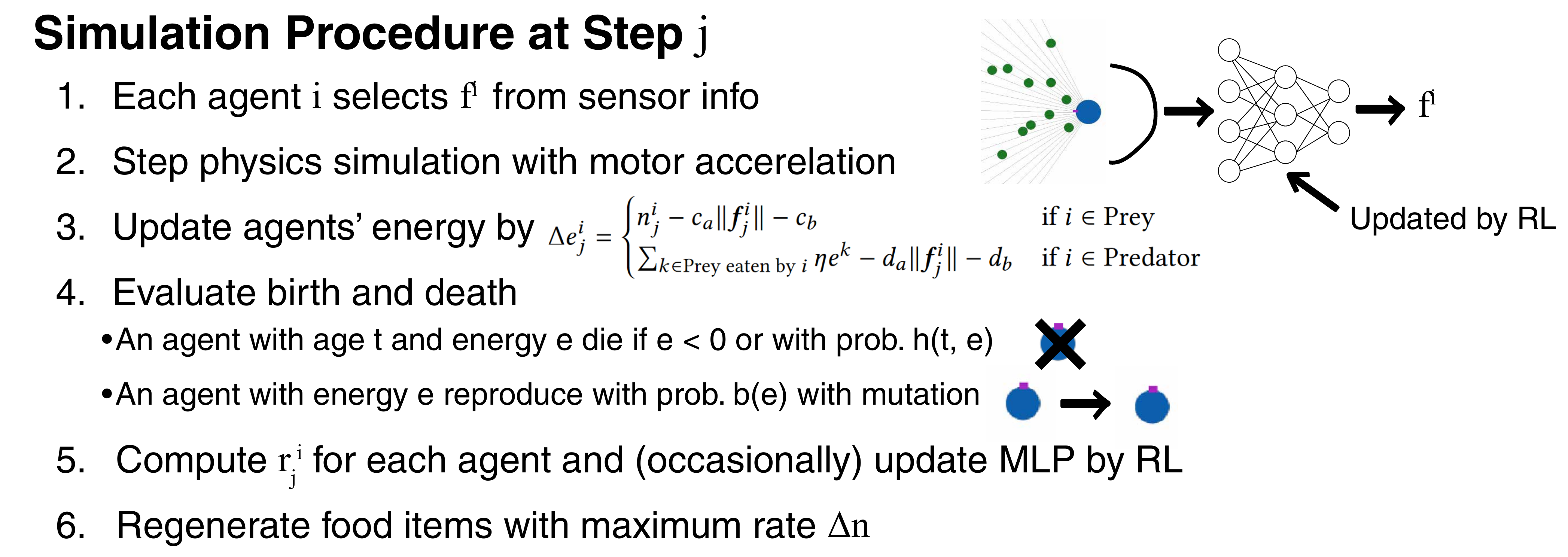}
  \caption{Simulation procedure.}~\label{figure:simulation}
\end{figure}

\Cref{figure:simulation} illustrates the simulation procedure. Each agent samples motor actions from an MLP, and the physical state is updated based on the resulting acceleration. Subsequently, the agents' energy levels are updated, followed by the birth and death processes.

\section{Results}\label{sec:results}
\begin{figure}[t]
  \centering
  \includegraphics[width=8cm]{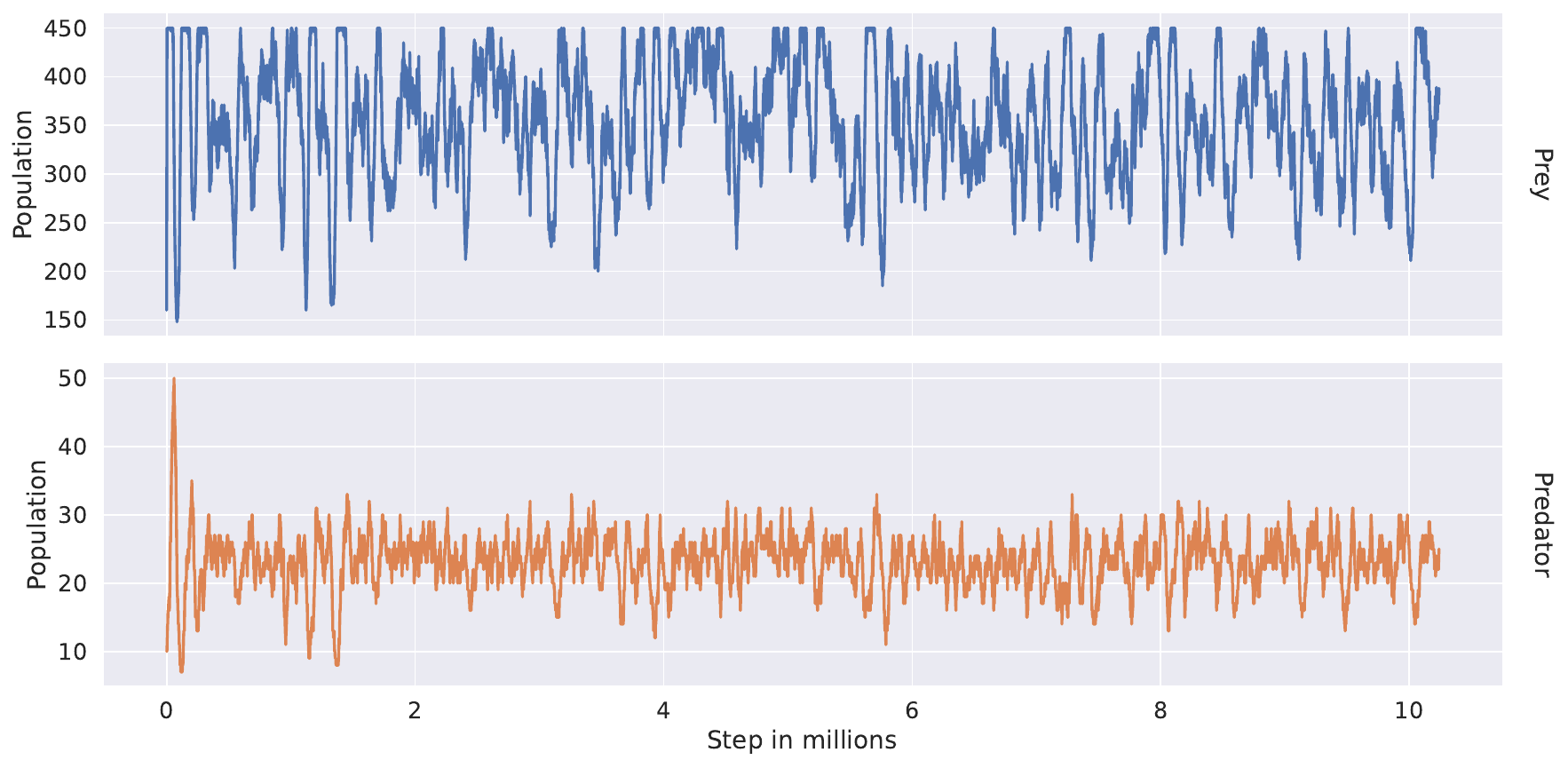}
  \includegraphics[width=8cm]{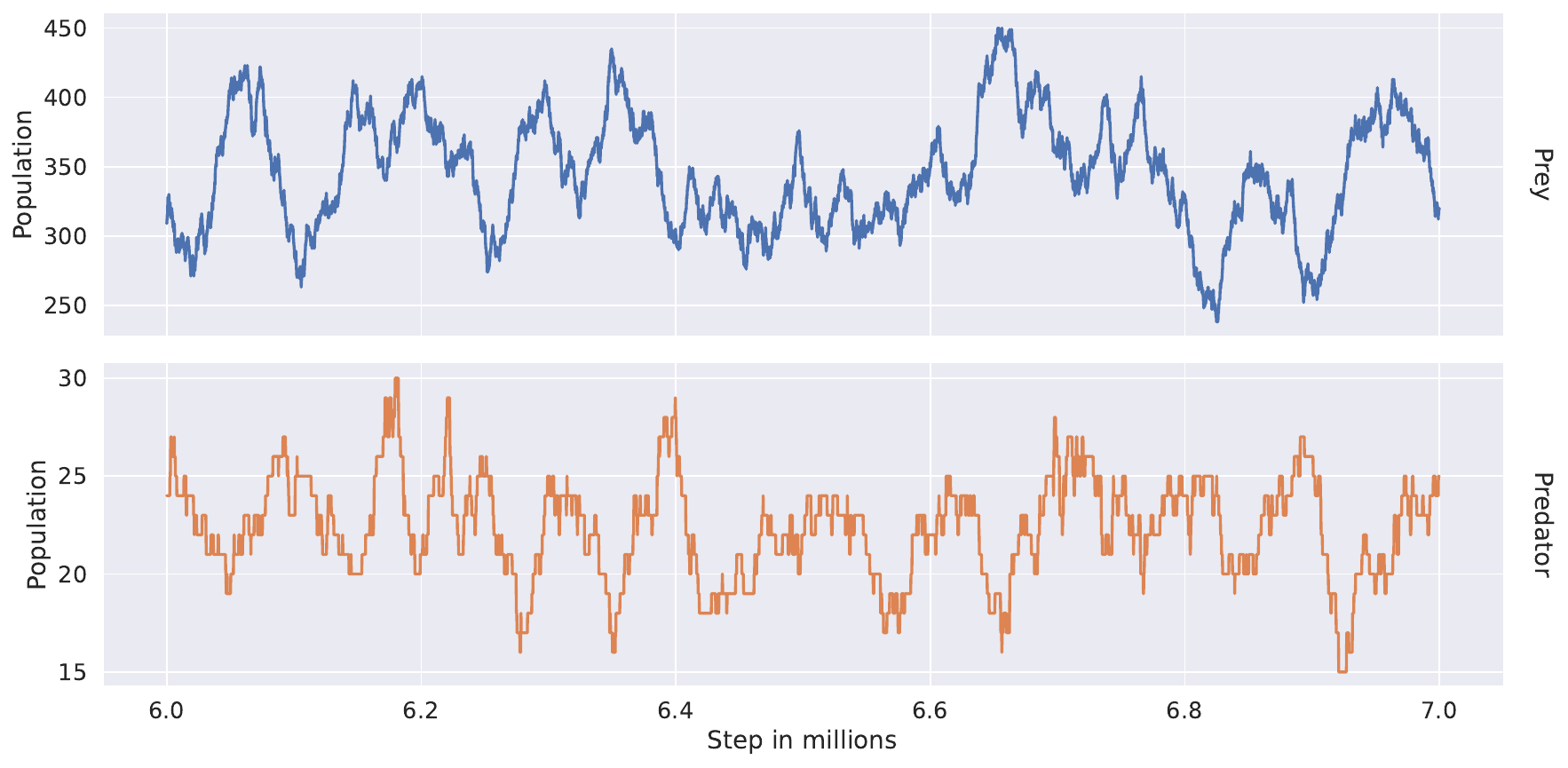}
  \caption{
    Fluctuations in prey and predator populations. Blue and orange lines represent prey and predator populations, respectively. Upper: The entire simulation timeline. Lower: A detailed view of the dynamics between $6 \times 10^6$ and $7 \times 10^6$ steps.
  }\label{figure:pop}
\end{figure}
This section begins with an analysis of evolved rewards under the default configuration, where predators possess medium-sized mouths and a food production rate $\Delta n = 0.5$. This is followed by a comparative analysis across various environmental conditions.
Finally, to evaluate how predator movement and lethality influence reward weights, we ran simulations in environments with stationary pitfalls that inflict critical damage to agents.

We conducted experiments totaling $10.24$ million steps, utilizing five distinct random seeds for each setting. The initial populations are set to $150$ prey and $10$ predators. Reward weights are sampled from a Gaussian distribution with a mean of $0$ and a standard deviation of $0.1$. These simulations required approximately $40$ hours of computation on a cluster node equipped with an NVIDIA A100 GPU. The anonymized codebase for these experiments is available for review and will be published upon the paper's release\footnote{\url{https://anonymous.4open.science/r/emevo-65EF}}.

Under default conditions, prey generation counts ranged from $473$ to $501$, while predator generation counts ranged from $47$ to $59$. \Cref{figure:pop} shows the population fluctuations in a simulation run with seed $1$, which shows that the oscillation occurs with a period of roughly one million steps.
The prey population increases when the predator population is low and declines as the predator population increases, a pattern qualitatively similar to that of the Lotka-Volterra model \citep{wangerskyLotkaVolterraPopulationModels1978}. However, unlike idealized mathematical models, the magnitudes of these oscillations exhibit substantial variability over the duration of the simulation.

\subsection{Analysis of Evolved Rewards}\label{sec:result-1}

\begin{figure*}
  \centering
  \includegraphics[width=\linewidth]{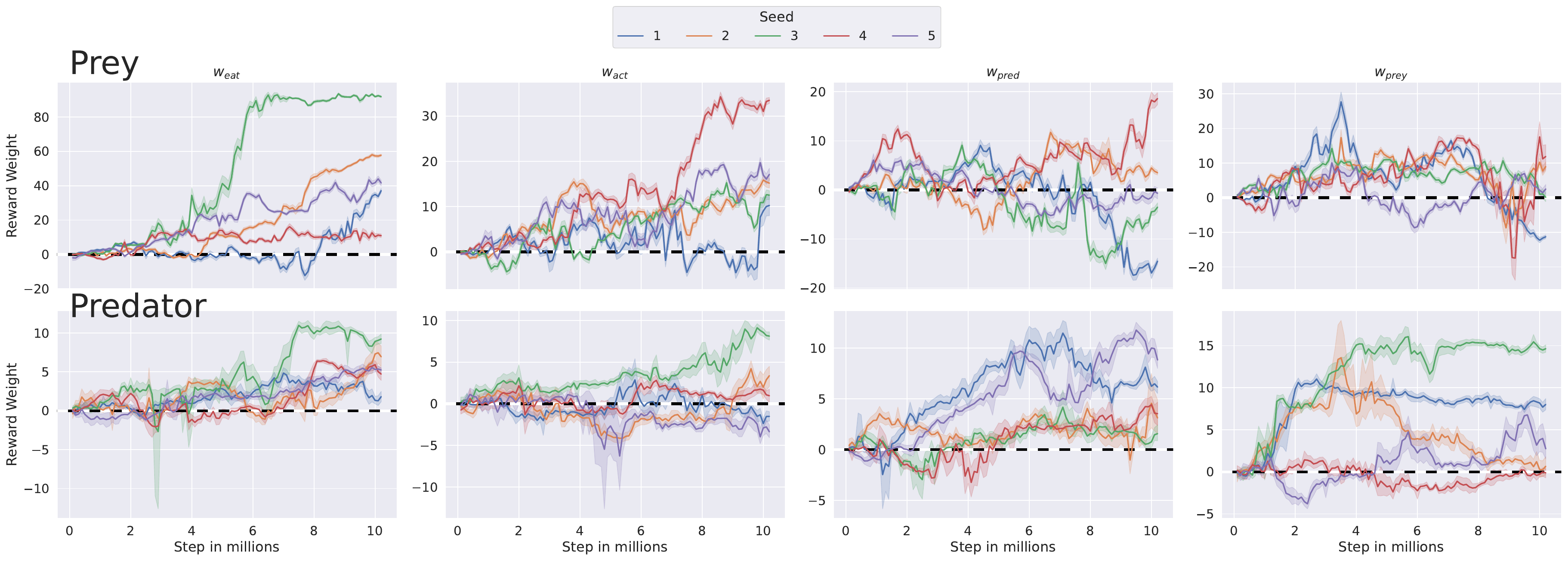}
  \caption{Evolution of reward weights $w_{\mathrm{eat}}$, $w_{\mathrm{act}}$, $w_{\mathrm{prey}}$, and $w_{\mathrm{pred}}$ (from left to right) for prey (top) and predators (bottom). Colors indicate results with different random seeds.}\label{figure:cum}
\end{figure*}

\Cref{figure:cum} shows the evolution of the reward weights over time, revealing distinct behavioral strategies emerging within the two populations. $w_\mathrm{eat}$ evolved toward significantly positive values for both prey and predators, reflecting the fundamental importance of food intake. Contrary, the evolution of $w_\mathrm{act}$ exhibited a notable divergence between the species. In the prey population, $w_\mathrm{act}$ remained consistently positive, suggesting a selective advantage for continuous movement or exploration. Conversely, the $w_\mathrm{act}$ of predators evolved either positive or negative values in different simulation runs.
The predators have fewer opportunities and face greater difficulties in hunting, which increases the need to conserve energy.

The evolution of predator rewards ($w_\mathrm{pred}$) and social rewards ($w_\mathrm{prey}$) within the prey population revealed a complex interplay between individual and collective survival strategies. While $w_\mathrm{pred}$ evolved either positively or negatively, $w_\mathrm{prey}$ evolved toward positive values in the majority of simulations. Notably, in the specific instances where $w_\mathrm{pred}$ evolved to be counter-intuitively positive (seeds $2$ and $4$), there was a concurrent evolution of strongly positive social rewards. This observation leads to the hypothesis that the emergence of sociality may have superseded the functional necessity for individual predator avoidance. We will confirm this hypothesis by analyzing the branching patterns within a single colony.

Social rewards for predators ($w_\mathrm{pred}$) evolved almost positively, which is reasonable because following another predator can guide a predator to regions with higher prey density.  Although $w_\mathrm{prey}$ reached positive values in most instances, it remained near zero in seeds 2 and 4, where the majority of prey agents exhibited positive $w_\mathrm{pred}$ values. The prey's attraction to predators may have reduced the evolutionary pressure on predators to rely on visual preference for prey.

\begin{figure}[t]
  \centering
  \includegraphics[width=\linewidth]{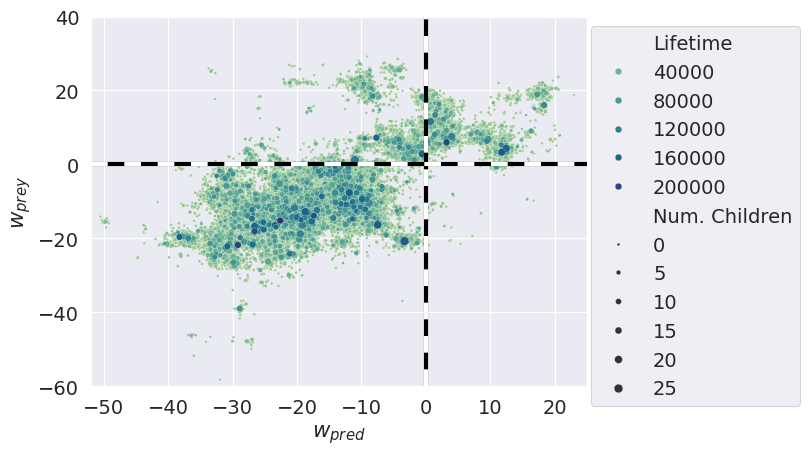}
  \caption{
    Scatter plot of the evolved rewards of prey agents born after 9 million steps in seed 1. Color represents the agent's lifetime, while dot size scales with the total number of offspring produced.
  }\label{figure:scatter}
\end{figure}

To visualize the variety of reward weights in a colony, we show the scatter plot of prey agents born after $9$ million steps in the experiment with seed $1$ in \Cref{figure:scatter}, in which two clusters are observed. A large portion of the agents is concentrated within the range $-40 < w_\mathrm{pred} < 0$ and $-30 < w_\mathrm{prey} < 0$. Another cluster with positive $w_\mathrm{prey}$ and both positive and negative $\mathrm{pred}$ values is also observed.

\begin{figure}[t]
  \centering
  \includegraphics[width=\linewidth]{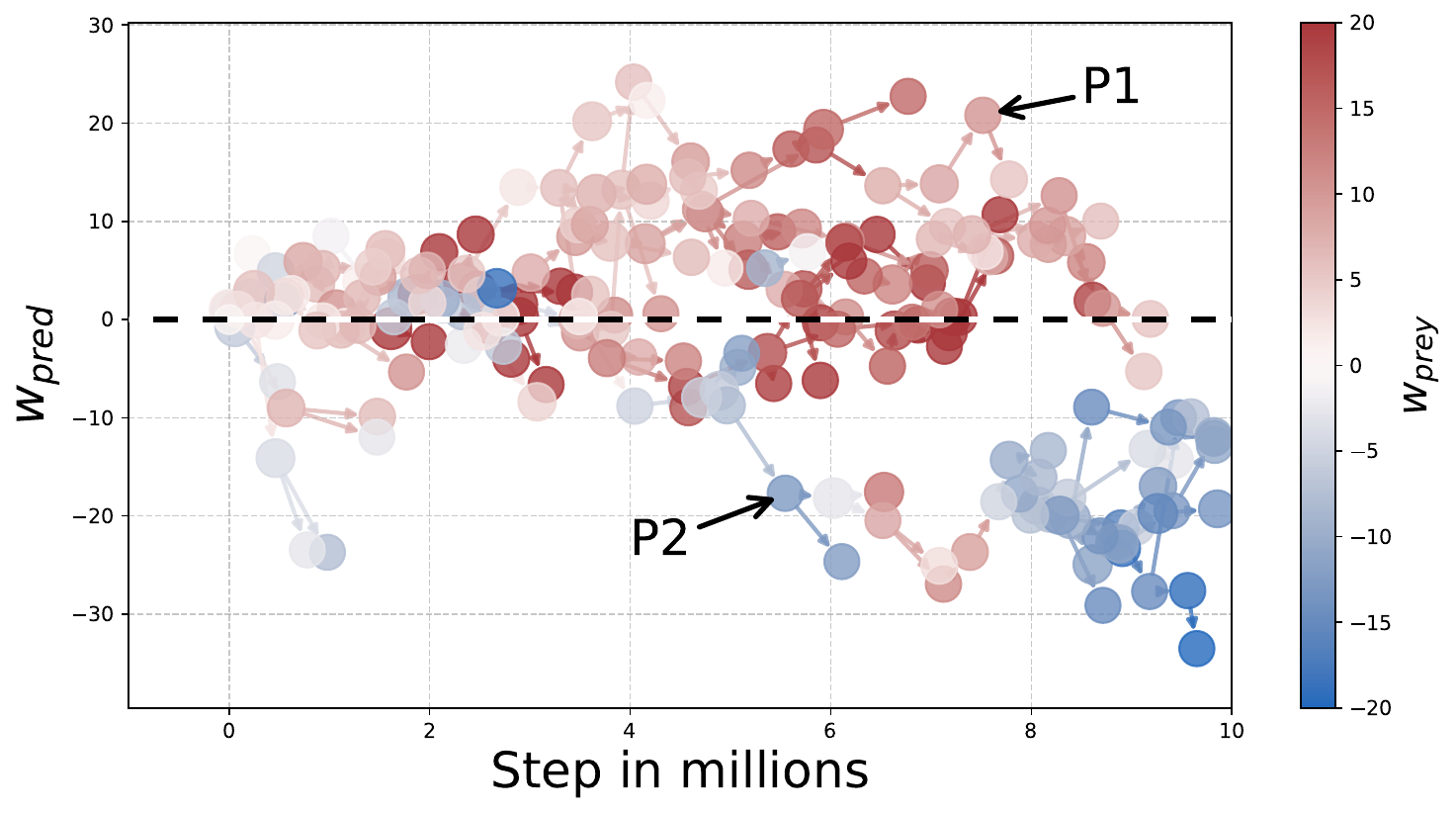}
  \caption{Compressed phylogenetic tree of all prey agents in seed $1$ shown in $(step \times w_\mathrm{pred})$ space colored by $w_\mathrm{prey}$.}\label{figure:tree}
\end{figure}

\begin{figure}[t]
  \centering
  \includegraphics[width=\linewidth]{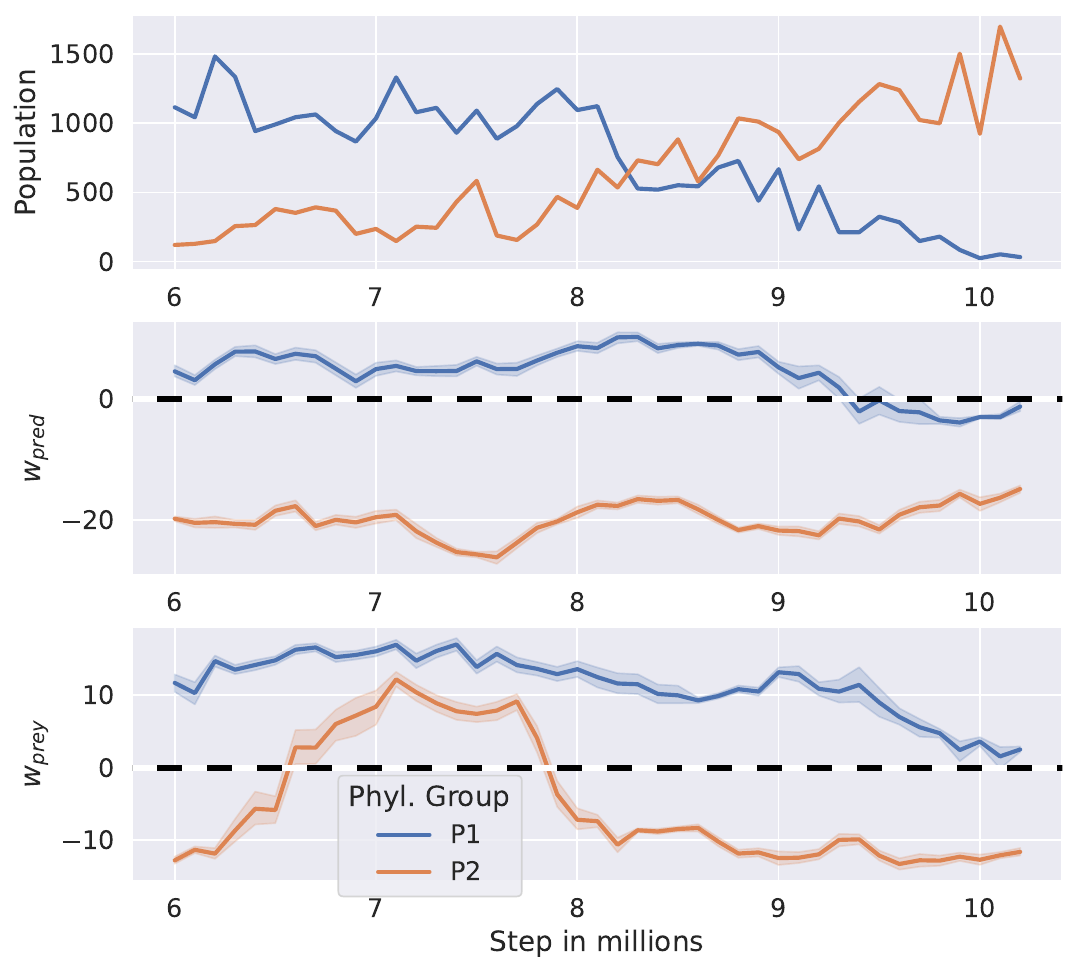}
  \caption{Population and the evolution of average $w_\mathrm{pred}$ and $w_\mathrm{prey}$ in prey agents in two phylogenetic groups P1 and P2 in \Cref{figure:tree}.}\label{figure:group-pop}
\end{figure}

To investigate why such branching into two phylogenetic groups happened, we show the phylogenetic tree of prey agents in \Cref{figure:tree}. In this visualization, the x-axis represents the simulation steps, the y-axis indicates the evolution of $w_\mathrm{pred}$, and the node colors signify the values of $w_\mathrm{prey}$. For visual clarity, the phylogenetic tree is compressed so that each node represents approximately $1000$ individuals, as described in \Cref{appendix:analysis}. Branching into two distinct groups is observed at approximately $5$ million steps. For further analysis, we label the group with mainly positive $w_\mathrm{pred}$ as P1 and the remaining population as P2.

\Cref{figure:group-pop} illustrates the population dynamics and the concurrent changes in $w_\mathrm{pred}$ and $w_\mathrm{prey}$ for both groups after $6$ million steps. Notably, the population of P2 eventually surpassed that of P1 between $8$ and $9$ million steps, indicating that P2 gained an evolutionary advantage over P1.

\begin{figure}[t]
  \centering
  \includegraphics[width=\linewidth]{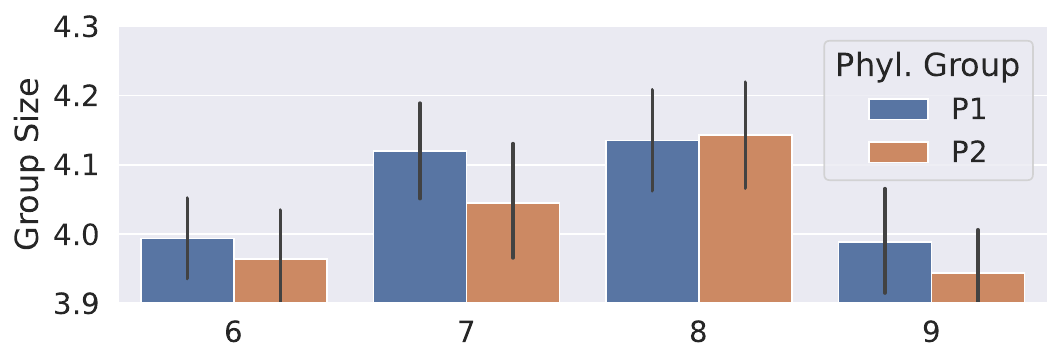}
  \includegraphics[width=\linewidth]{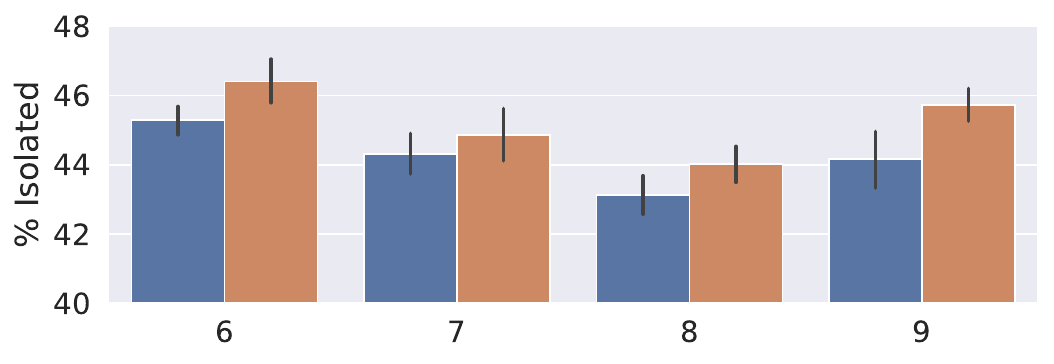}
  \includegraphics[width=\linewidth]{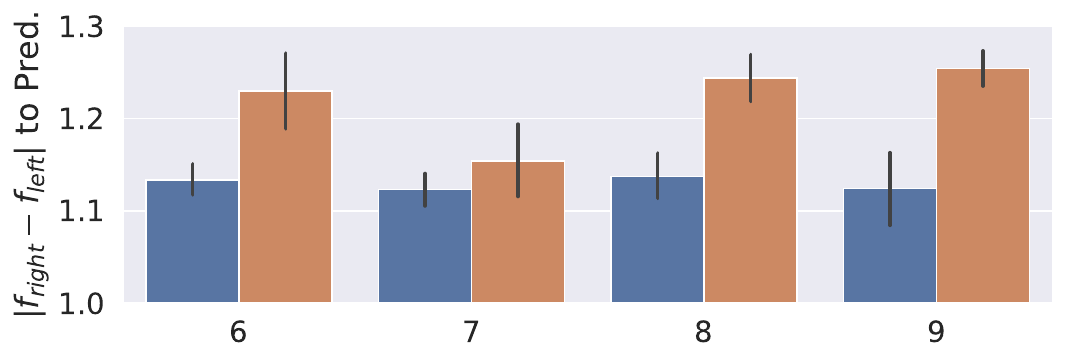}
  \caption{
    Behavior statistics of prey agents in phylogenetic groups P1 (blue) and P2 (orange) in \Cref{figure:tree}.
    The panels display the average group size of prey agents (top), the percentage of isolated prey agents (middle), and the average $|f_{right} - f_{left}|$ in the faces of predators (bottom).}\label{figure:group-bars}
\end{figure}

To examine the influence of $w_\mathrm{pred}$ and $w_\mathrm{prey}$ on collective and escaping behaviors, we evaluate the average group sizes, the percentage of isolated prey, and the reaction of prey agents to predators for groups P1 and P2 from $6$ million steps onward in \Cref{figure:group-bars}. We quantified the prey agent's response to a predator by averaging $|f_{right} - f_{left}|$ when the predator is observed at close range, which represents turning away from the predator. Further details regarding the analysis are provided in \Cref{appendix:analysis}. The results show that prey agents in P2 tend to belong to smaller groups and are more isolated, as indicated by negative $w_\mathrm{prey}$. While phylogenetic group P1 maintains larger group sizes, 
prey agents in P2 tend to exhibit stronger avoidance in the face of predators. These results indicate that two distinct survival strategies, learned through distinct reward functions, coexist. P1 primarily relies on social grouping as a defense, while P2 relies more on individual escape behavior.

While this analysis focused on seed $1$, where a majority of the agents evolved to have fear with negative $w_\mathrm{pred}$, similar branching patterns were observed in other seeds. Examples include seed $2$ and $4$, where a positive $w_\mathrm{pred}$ was the dominant trait across the population, as we show in \Cref{appendix:results}. The consistent appearance of these lineages suggests that, in environments where both collective and solitary survival strategies are effective, they can coexist within the same population. Furthermore, it appears that counter-strategies that contrast with the majority can be effective, enhancing the branching of reward weights.

\subsection{Predators with Different Mouth Sizes}\label{sec:result-mouth}
\begin{table}[t]
  \centering
  \begin{tabular}{lcc}
    \toprule
    Setting & Prey Pop. & Predator Pop.\\
    \midrule
    Default (M. Mouth and $\Delta n = 0.5$) & $349.26$ & $23.09$ \\
    Small Mouth & $398.21$ & $23.48$ \\
    Large Mouth & $331.28$ & $22.21$ \\
    Less Food ($\Delta n = 0.4$) & $374.76$ & $18.50$ \\
    More Food ($\Delta n = 0.6$) & $342.26$ & $25.41$ \\
    Pitfall & $424.36$ & $-$ \\
   \bottomrule
  \end{tabular}
  \caption{Average prey and predator population in all settings.}\label{table:pop}
\end{table}

To investigate how the hunting capability of predators influences the evolution of fear in prey, we conducted experiments using predators with three distinct mouth sizes (small (S), medium (M), and large (L)) shown in \Cref{figure:mouth}. Note that M is the default and used in the previous section. For the large-mouth-size condition, we conducted 6 simulation runs because the predator population went extinct in one random seed. This run was excluded from the results and analyzed in \Cref{appendix:results}. As shown in \Cref{table:pop}, an increase in predator mouth size correlates with a reduction in the prey population, whereas the predator population remains largely unaffected.

\begin{figure*}[t]
  \includegraphics[width=16cm]{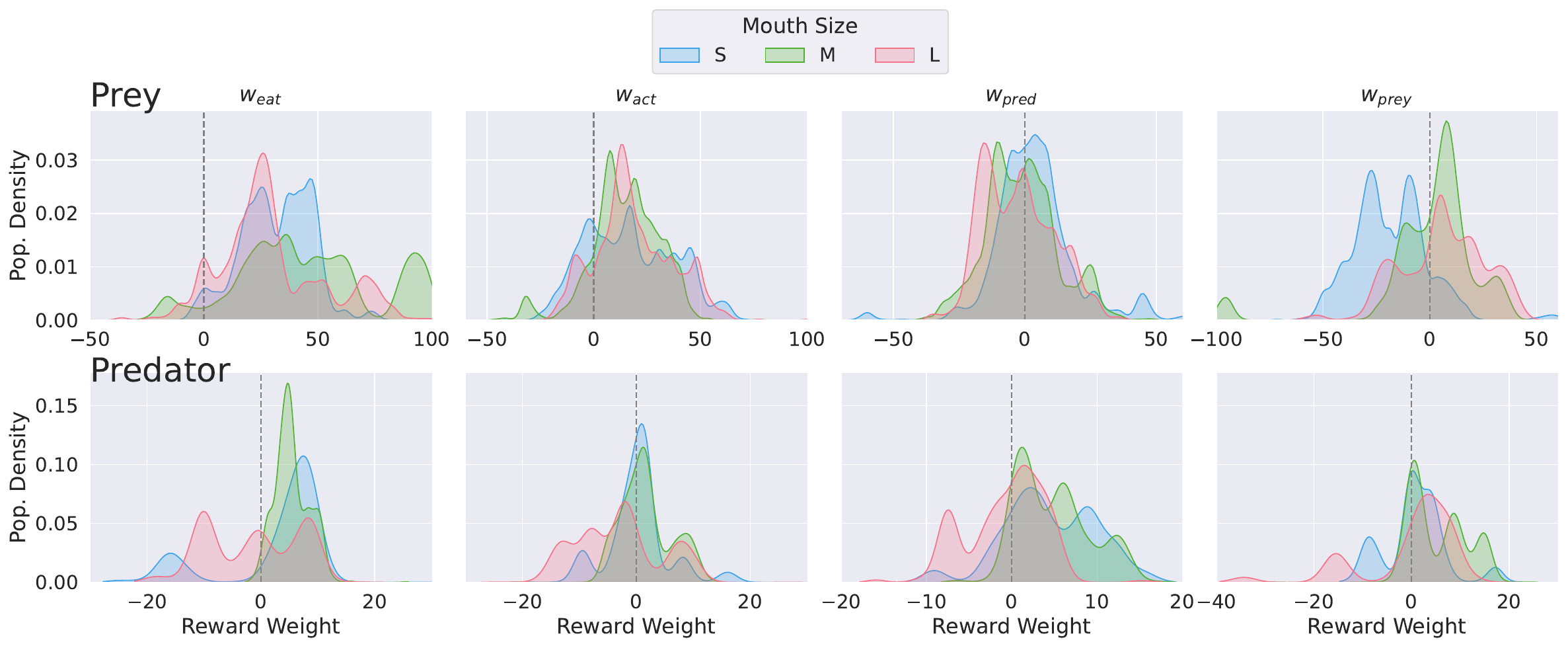}
  \caption{Comparison of evolved reward weights of prey (top) and predators (bottom) between environments with different predator mouth sizes. Blue corresponds to small mouths, green to medium, and red to large. }\label{figure:kde-mouth}
\end{figure*}

We compare the evolved reward weights for prey and predators born after $9$ million steps with different predator mouth sizes in \Cref{figure:kde-mouth}.
We utilized Kernel Density Estimation (KDE) for this plot. The results indicate that $w_\mathrm{pred}$ is more negative and $w_\mathrm{prey}$ is larger for prey when the predator mouth size is larger, suggesting that against stronger predators, both fear and social rewards are more important. On the other hand, $w_\mathrm{eat}$ is smaller, implying more importance of fear and social rewards compared to food intake.

\begin{figure}[t]
  \centering
  \includegraphics[width=\linewidth]{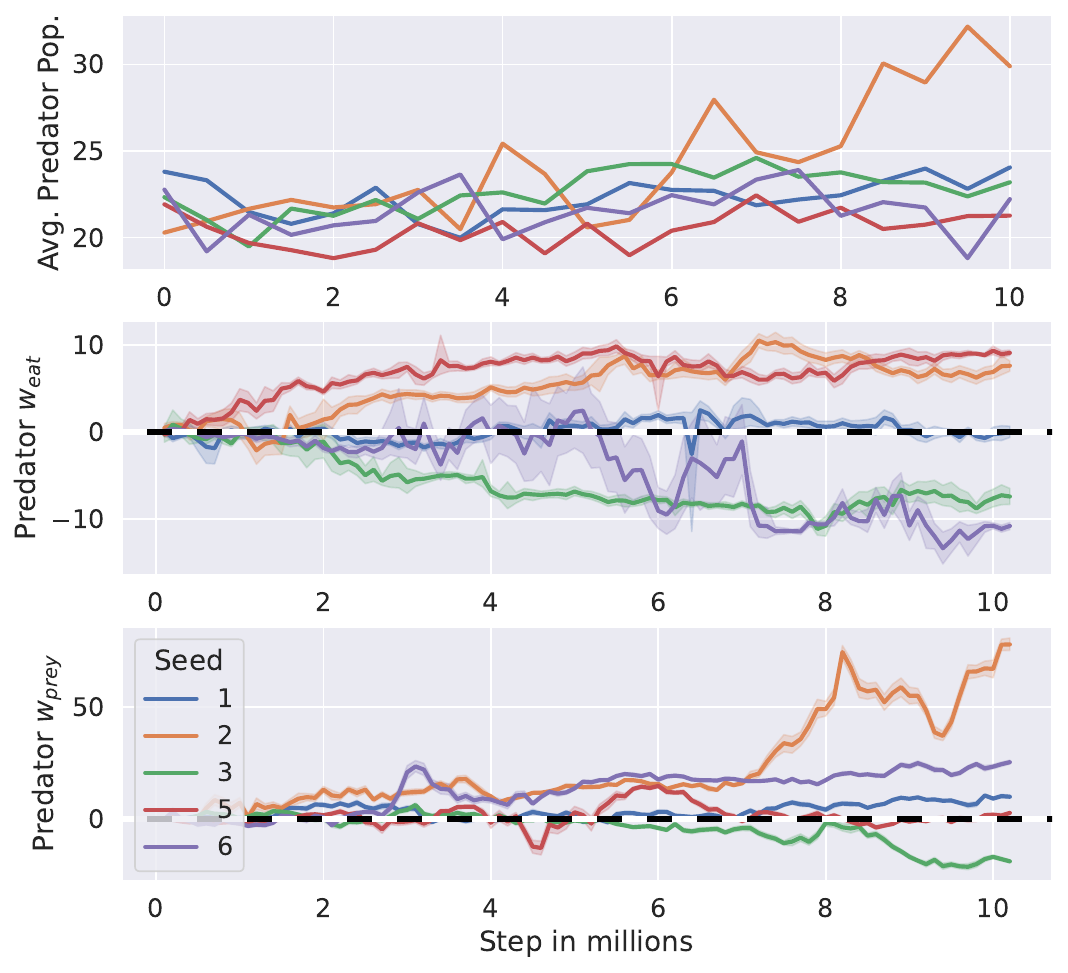}
  \caption{Predator population averaged over $0.5$ million steps, average $w_\mathrm{eat}$, and average $w_\mathrm{prey}$ per seed in experiments with the large predator mouth.}\label{figure:predator-l}
\end{figure}

Regarding the evolved reward weights of the predators, we found that $w_\mathrm{eat}$ is notably larger when they possess smaller mouths. This likely occurs because predators with limited hunting capabilities have fewer opportunities to feed, making the food reward more crucial for survival. Surprisingly, $w_\mathrm{eat}$ even becomes negative when the mouth size is large. \Cref{figure:predator-l} shows the predator population (averaged over $0.5$ million steps), $w_\mathrm{eat}$, and $w_\mathrm{prey}$ of predators with large mouths for five runs. From this figure, only seed 2, which possesses positive $w_\mathrm{eat}$ and $w_\mathrm{prey}$, exhibited a remarkable population increase. However, as we analyze in \Cref{appendix:results}, being too greedy is risky and can lead to extinction in this environment by abruptly reducing the prey population. Thus, while positive $w_\mathrm{eat}$ and $w_\mathrm{prey}$ certainly contribute to the population increase of predators, when their mouth size is large enough, a negative $w_\mathrm{eat}$ may contribute to sustainable hunting to maintain population balance.

\begin{figure*}[t]
  \includegraphics[width=16cm]{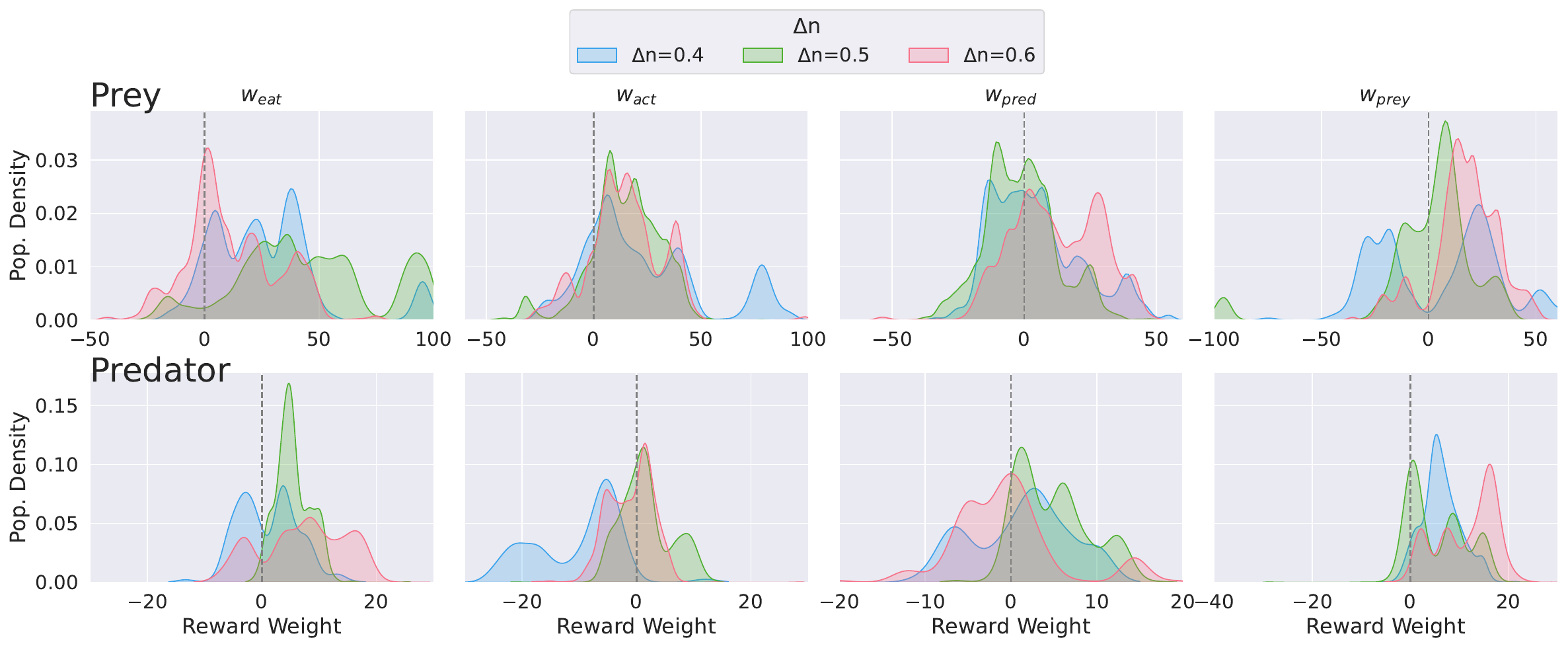}
  \caption{Comparison of evolved reward weights of prey (top) and predators (bottom) between environments with different food regeneration ratios $\Delta n$. Blue corresponds to $\Delta n = 0.4$, green to $\Delta n = 0.5$ (default), and red to $\Delta n = 0.6$.}\label{figure:kde-delta-n}
\end{figure*}

\subsection{Resource Richness}
To examine the effect of environmental richness on the evolution of fear and social rewards, we conducted experiments with varying food regeneration ratios $\Delta n$. We tested $\Delta n = 0.4, 0.5$, and $0.6$, with $\Delta n = 0.5$ serving as the baseline result described in \Cref{sec:result-1}. Due to the higher frequency of extinction events, we utilized $5$ out of $6$ runs for $\Delta n = 0.4$ and $5$ out of $13$ runs for $\Delta n = 0.6$. As shown in \Cref{table:pop}, we observed a surprising result: the prey population decreases as $\Delta n$ increases, whereas the predator population increases with $\Delta n$. This suggests that resource richness affects the predator population more than the prey population.

We compared the evolved reward weights of agents born after $9$ million steps by KDE plot in \Cref{figure:kde-delta-n}. In the rich environment ($\Delta n = 0.6$), a smaller proportion of prey agents exhibit fear, and the majority of them have positive $w_\mathrm{pred}$. However, the difference between the $\Delta n = 0.4$ and $\Delta n = 0.5$ cases is subtle. This may be explained by the significant decrease in the predator population. It's dropping from $23$ to $18$ when $\Delta n$ falls from $0.5$ to $0.4$, which reduces the evolutionary pressure to maintain fear.

As $\Delta n$ decreases, more prey agents prefer solitary behaviors, as the ratio of agents with negative $w_\mathrm{prey}$ increases. This is contrary to the results with varying predator mouth sizes, where more prey agents evolved positive social rewards as the environment became more severe. This suggests that, when food is scarce, being solitary for foraging is more important than collective grouping for avoiding predators.

In the severe environment ($\Delta n = 0.4$), the predators' $w_\mathrm{eat}$ and $w_\mathrm{act}$ are smaller, reflecting the need to save energy. Conversely, in the richest environment ($\Delta n = 0.6$), $w_\mathrm{eat}$ and $w_\mathrm{prey}$ are the largest among the three settings. This may reflect competition among predators, as this environment has fewer prey and more predator agents.

\subsection{Evolution of Fear to Pitfalls}

\begin{figure}[t]
  \centering
  \includegraphics[width=8cm]{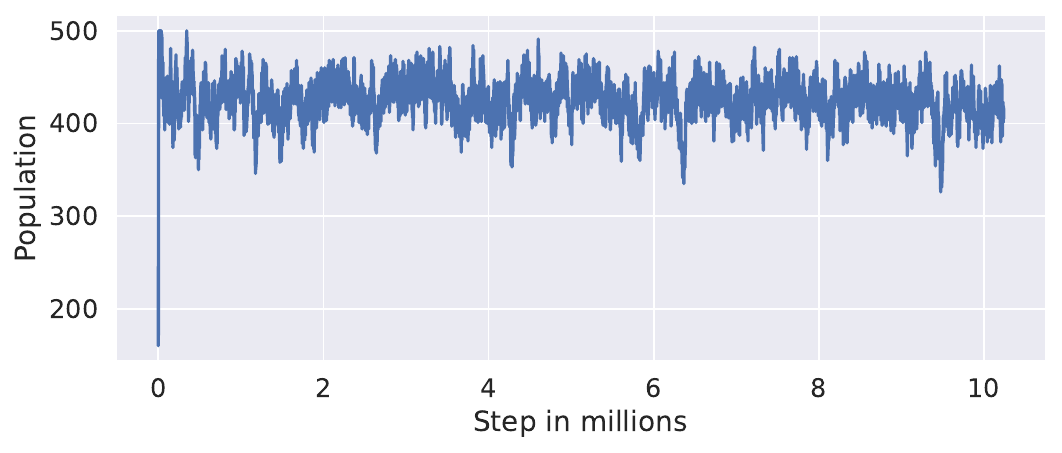}
  \caption{Population dynamics shown in the pitfall experiment with seed $4$.}\label{figure:pop-pit}
\end{figure}

Lastly, we conducted experiments in an environment with harmful pitfalls instead of predators. We conducted experiments with 7 random seeds and reported results for the 5 seeds in which the agents did not go extinct. \Cref{figure:pop-pit} illustrates the population change in this environment for seed $4$. Although the population still exhibits some oscillation, the magnitude of these fluctuations is smaller and the interval is shorter than those observed in the prey-predator experiments. This difference highlights the significant impact that predators have on driving broader population cycles.

\begin{figure*}[t]
  \includegraphics[width=16cm]{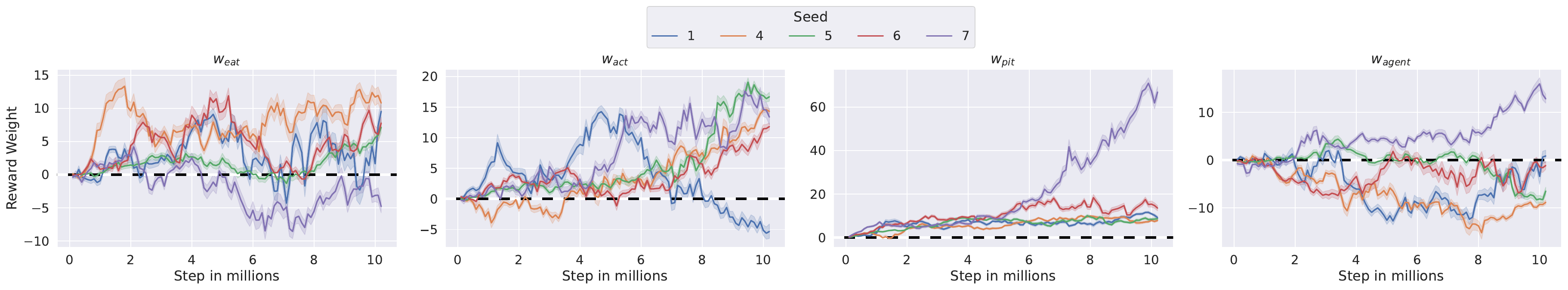}
  \caption{Evolution of reward weights $w_{\mathrm{eat}}$, $w_{\mathrm{act}}$, $w_{\mathrm{pit}}$, and $w_{\mathrm{agent}}$ (from left to right) in the pitfall environment. Colors indicate results with different random seeds.}\label{figure:cum-pit}
\end{figure*}

In \Cref{figure:cum-pit}, we plotted the evolution of reward weights $w_\mathrm{eat}$, $w_\mathrm{act}$, $w_\mathrm{pit}$ and $w_\mathrm{agent}$ per different random seeds. The results show that the weights for food ($w_\mathrm{eat}$) and action ($w_\mathrm{act}$) rewards remained largely positive. In contrast, the weight for other agents ($w_\mathrm{agent}$) was consistently negative, which suggests that the agents developed a preference for solitary behavior in this setting.

\begin{figure}[t]
  \includegraphics[width=6cm]{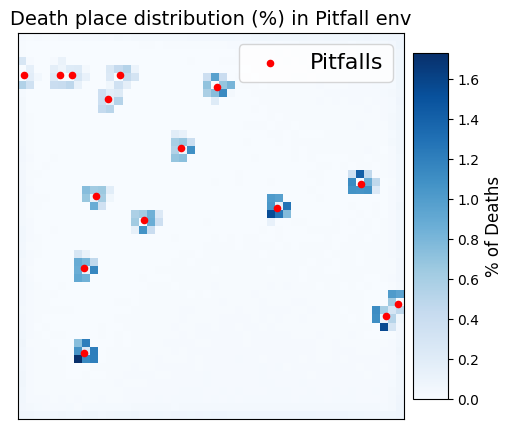}
  \caption{Death place distribution in the pitfall environment with seed $4$.}\label{figure:death-place}
\end{figure}

Surprisingly, $w_\mathrm{pit}$ evolved positively across all seeds. We suspect this trend emerged because food distribution is influenced by the location of static pitfalls. As shown in the snapshot from a representative evolution in \Cref{subfig:obsenv} (around $8$ million steps in seed $4$), many food items tend to cluster around pitfalls. Consequently, approaching pitfalls becomes a risky but reasonable strategy for agents to obtain sufficient food for reproduction. This is further supported by the distribution of death places in seed $4$ (\Cref{figure:death-place}), which indicates that many agents died in close proximity to pitfalls, underscoring the vital importance of retrieving food items from these dangerous areas.

\begin{figure}[t]
  \includegraphics[width=6cm]{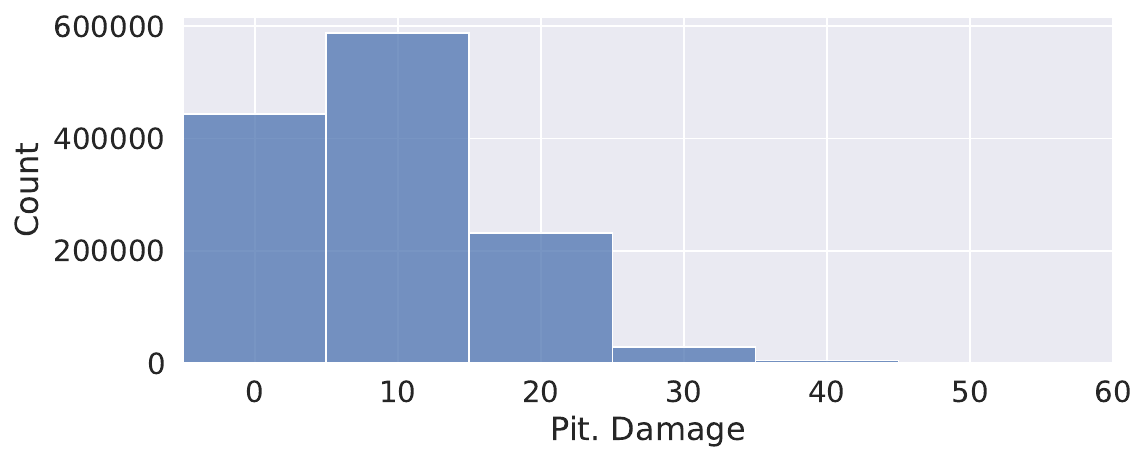}
  \caption{Histogram of the total pitfall damage sustained by agents evolved in the pitfall environment.}\label{figure:damage-count}
\end{figure}

Another distinguishing feature of the pitfall environment is that the $-10$ energy penalty is not lethal for agents with high foraging productivity. This is reflected in \Cref{figure:damage-count}, which displays the histogram of total pitfall damage across all agents. The data indicate that some individuals managed to survive despite colliding with pitfalls three or four times. This lack of lethality can lead to the evolution of a preference for pitfalls.

These findings contrast with the prey-predator simulations, in which a distinct fear of the predator evolved in the majority of prey agents. The static nature of the pitfalls led to an unbalanced food distribution, creating a different type of survival pressure. Furthermore, because the number of pitfalls remains constant, unlike the fluctuating predator populations, the survival pressure is stable. This lack of hunting movement, lethality, and population growth likely makes the pressure from pitfalls significantly weaker than that of predators, allowing agents to prioritize food acquisition even in dangerous areas.

\section{Conclusion}
This study investigated the evolutionary relationship between fear, social rewards, and predation. Our simulations demonstrate that while fear of predators generally evolves,
positive reward for predators can evolve together with social reward for prey, and weak prey reward for predators.
A pattern was also observed in which a group with high social reward and low fear of predators coexisted with, and was later replaced by another group with higher fear. We found that although increasing a predator’s hunting proficiency significantly amplified the evolution of fear, environments with high resource severity did not promote fear, thereby restricting the predator population. Finally, fear did not emerge in response to static pitfalls, suggesting that the predator's hunting movements, lethality, and dynamic proliferation amplify the evolutionary need for fear.

Our results highlight the importance of predators in the evolution of fear and group reward. These findings demonstrate that prey-predator interactions can drive the evolution of diverse and complex reward structures. Furthermore, the findings suggest that social situations can drive the evolution of emotions and motivations beyond basic fear or familiarity. Our results also showed the possibility of multi-agent evolutionary simulations for explaining diverse animal behaviors beyond optimal decision-making that AI research typically focuses on.

\clearpage
\bibliographystyle{apalike}
\bibliography{references}

\clearpage
\appendix
\section{Parameters used in experiments}\label[apsec]{appendix:param}

The simulation environment is a square domain measuring $960 \times 960$ units. Within this space, agents are modeled as circular entities: the prey has a radius of $10$ units, while the predator has a radius of $14$ units. To detect environmental features and other agents, they are equipped with proximity sensors with a maximum length of $120$ units.

\begin{table}[t]
  \centering
  \begin{tabular}{ccc}
    \toprule
    Parameter & Value \\
    \midrule
    $c_b$ & $2.5e-6$ \\
    $c_a$ & $1e-4$ \\
    $d_b$ & $4e-3$ \\
    $d_a$ & $5e-5$ \\
    \bottomrule
  \end{tabular}
  \caption{Energy consumption parameters used in our experiments.}\label{table:ec}
\end{table}

\Cref{table:ec} shows energy consumption parameters used in our experiments. Because $0 < \|\mathbf{f}\| < 114$, $c_a\|\mathbf{a}\|_j$ is twice as large as $c_b$ when the motor output is the maximum. Also, predators consume about $10$ times as much energy as prey.

\begin{table}[t]
  \centering
  \begin{tabular}{cc}
    \toprule
    Parameter & Value \\
    \midrule
    $\kappa_{h}$ & 0.01 \\
    $\alpha_{e}$ & 0.02 \\
    $\beta_{h}$ & 0.2 \\
    $\alpha_{t}$ & \num{4e-7} (prey) / \num{2e-7} (predator) \\
    $\beta_{t}$ & \num{2e-6} (prey) / \num{4e-6} (predator) \\
    $\zeta$ & 10 (prey) / 100 (predator) \\
    $\kappa_{b}$ & \num{1e-3} \\
    $\beta_{b}$ & 0.1 \\
    \bottomrule
  \end{tabular}
  \caption{Parameters used for hazard function $h$ and birth function $b$, which are common to both prey and predators unless otherwise noted.}\label{table:hb}
\end{table}

\Cref{table:hb} shows the parameters used in the birth and hazard models in our experiments. These parameters are tuned so that both prey and predators can maintain sufficient populations for evolution, with a subtle change in the predator's mouth size and $\Delta n$.

\begin{figure}[t]
\includegraphics[width=7cm]{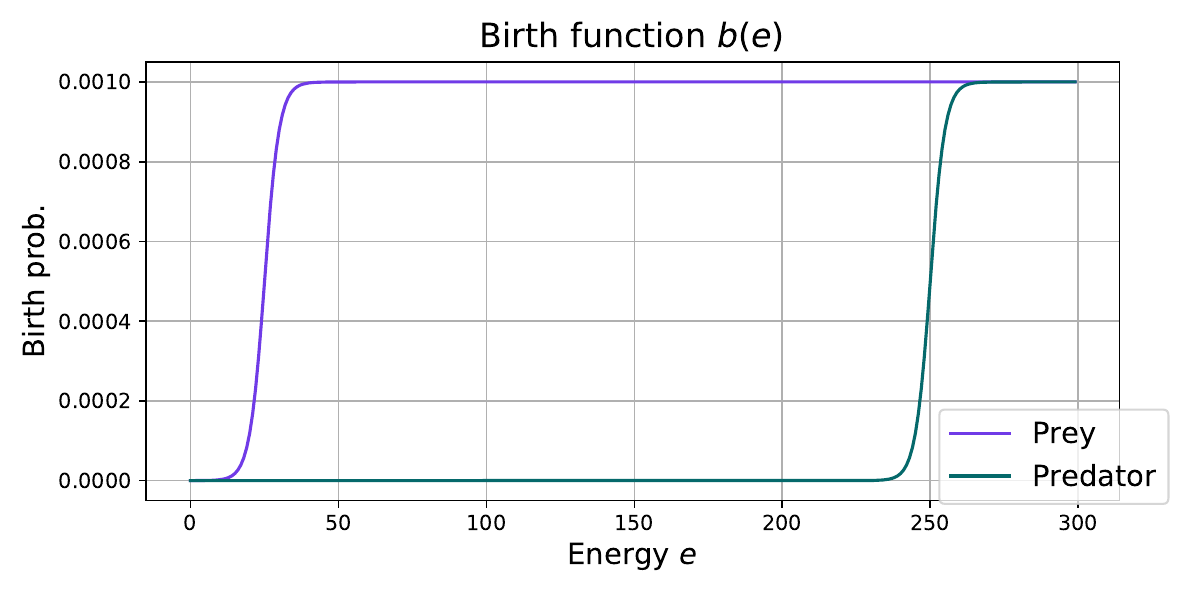}
\caption{Birth for prey (purple) and predators (green).}\label{birth}
\end{figure}

We plot the birth function of prey and predator agents with the given parameters in \Cref{birth}. We can see that $30$ energy units are required to increase the birth probability for prey agents, while predators need $260$ or more.

\begin{figure}[t]
  \centering
  \includegraphics[width=8cm]{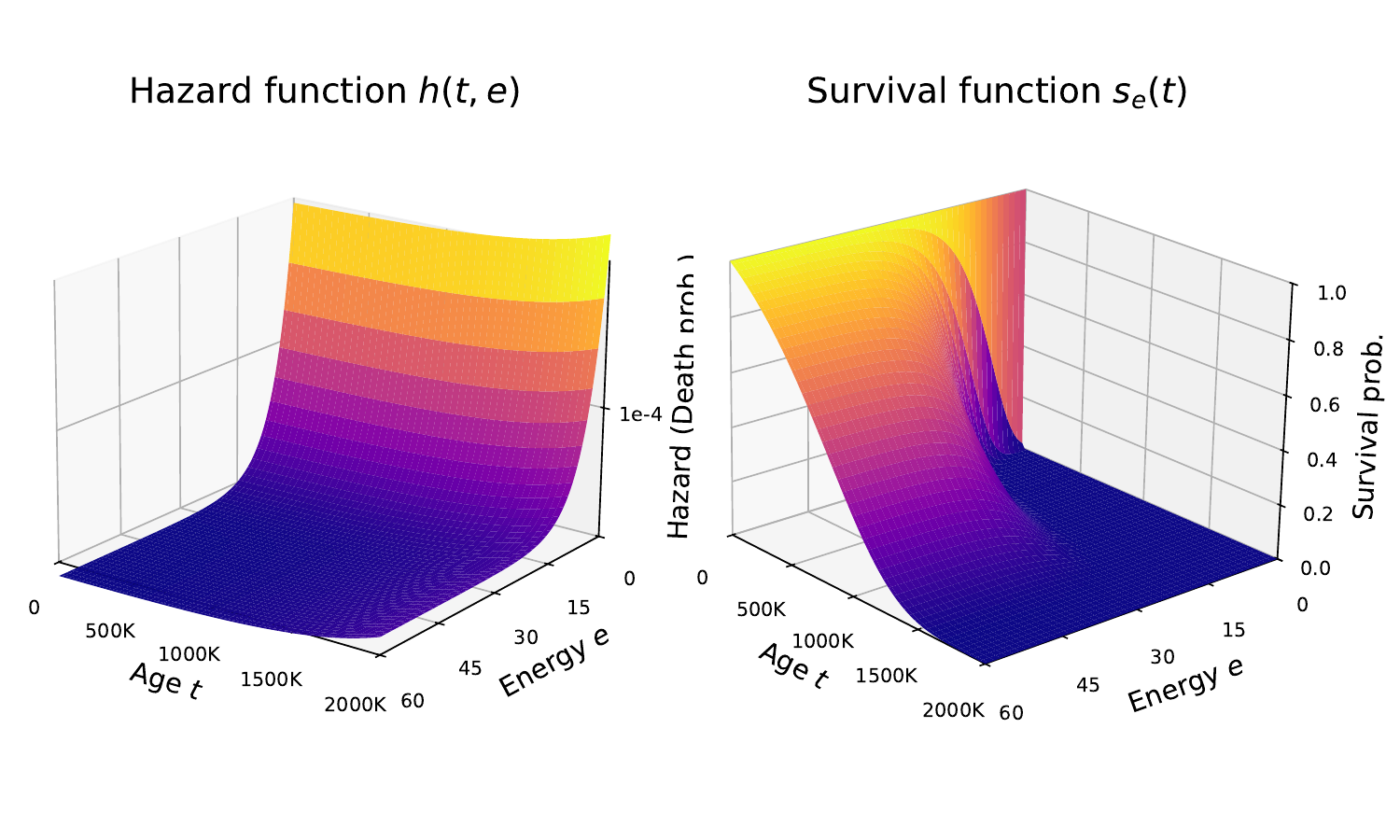}
  \caption{Hazard (left) and survival function (right) for prey. We can see that the survival probability significantly decreases when $e < 20$ while the effect of aging is milder.}\label{hazard}
\end{figure}

We plot the hazard function and the survival function $S_{e}(t) = \exp \left( -\int_{0}^{t} h(t, e)dt \right)$ derived from it only for prey agents. We can see that the death probability increases suddenly when the energy level falls under $15$.

\begin{table}[t]
  \centering
  \begin{tabular}{ll}
    \toprule
    Parameter & Value \\
    \midrule
    Discount factor ($\gamma$) & $0.999$ \\
    Rollout steps ($N$) & $1024$ \\
    Minibatch size & $256$ \\
    Number of optimization epochs & $10$ \\
    PPO Clipping parameter & $0.2$ \\
    Entropy coeff. & $\num{0.001}$ \\
    GAE parameter ($\gamma$) & $0.95$ \\
    Adam learning rate & $\num{3e-4}$ \\
    Adam $\epsilon$ & $\num{1e-7}$ \\
    Size of hidden layer in MLP & $64$ \\
    \bottomrule
  \end{tabular}
  \caption{RL parameters}\label{table:rl-param}
\end{table}

We show the RL parameters used in \Cref{table:rl-param}, tuned based on those used in the continuous control experiments in the PPO paper~\citep{schulmanProximalPolicyOptimization2017}.

In the initial population, reward weights ($w_{\mathrm{eat}}$, $w_{\mathrm{act}}$, $w_{\mathrm{prey}}$, and $w_{\mathrm{pred}}$) is sampled from a normal distribution with center $0.0$ and standard deviation $0.1$.

\section{Analysis Details}\label[apsec]{appendix:analysis}

\subsection{Phylogenetic Tree}
The phylogenetic tree shown in \Cref{figure:tree} is obtained by the following algorithm:

\begin{enumerate}
  \item Calculate the total number of descendants for every node in the original phylogenetic tree.
  \item Starting from an arbitrary leaf node, traverse upward toward the root to find the first node that possesses more than $1000$ descendants.
  \item Detach the identified node from the main tree and update the descendant counts for all remaining ancestral nodes to reflect the removal.
  \item Repeat the (2) and (3) until no further nodes meeting the descendant threshold can be identified.
  \item Construct the final compressed tree structure using the split nodes.
\end{enumerate}

\subsection{Group Analysis}
Phylogenetic group P2 in \Cref{figure:tree} is identified by locating the prey agent exhibiting the first significant jump in $w_\mathrm{pred}$ value when traversing from the root node of the indicated compressed group.

To calculate the group size in \Cref{figure:group-bars}, we took the following procedure at each sampled step:

\begin{enumerate}
    \item We excluded agents located near the boundaries ($x, y < 60$ or $x, y > 900$) to eliminate inactive groups that tend to aggregate along the walls.
    \item For the remaining prey agents, two individuals were defined as belonging to the same group if the Euclidean distance between them was less than $10$.
    \item The final group sizes were computed by identifying the connected components of the resulting proximity graph using the \texttt{scipy.sparse.csgraph.connected\_components} function\footnote{SciPy: \url{https://scipy.org/}, accessed January 17, 2026.}.
\end{enumerate}

If a resulting group size is $1$, the agent is classified as isolated. The total percentage of isolated agents within the population is then computed. For all agents in groups of size greater than 1, we calculate the average group size of each agent's cluster. Both statistics are averaged over all sampled steps (once every $1000$ steps) and presented as bar graphs.

The average  $f_{right} - f_{left}$ is also computed for all agents located away from the environment boundaries ($x, y \ge 60$ and $x, y \le 900$). An agent is defined a facing a predator if at least one predator is detected within the prey's proximity sensor range, specifically within a $120$ degree forward-facing arc, at a distance of less than $60$ units (representing half of the maximum sensor range).

\section{Additional Results}\label[apsec]{appendix:results}
\subsection{Variation of Phylogenetic Trees}
\begin{figure*}[t]
  \centering
  \begin{subfigure}{0.24\linewidth}
    \centering
    \includegraphics[width=\linewidth]{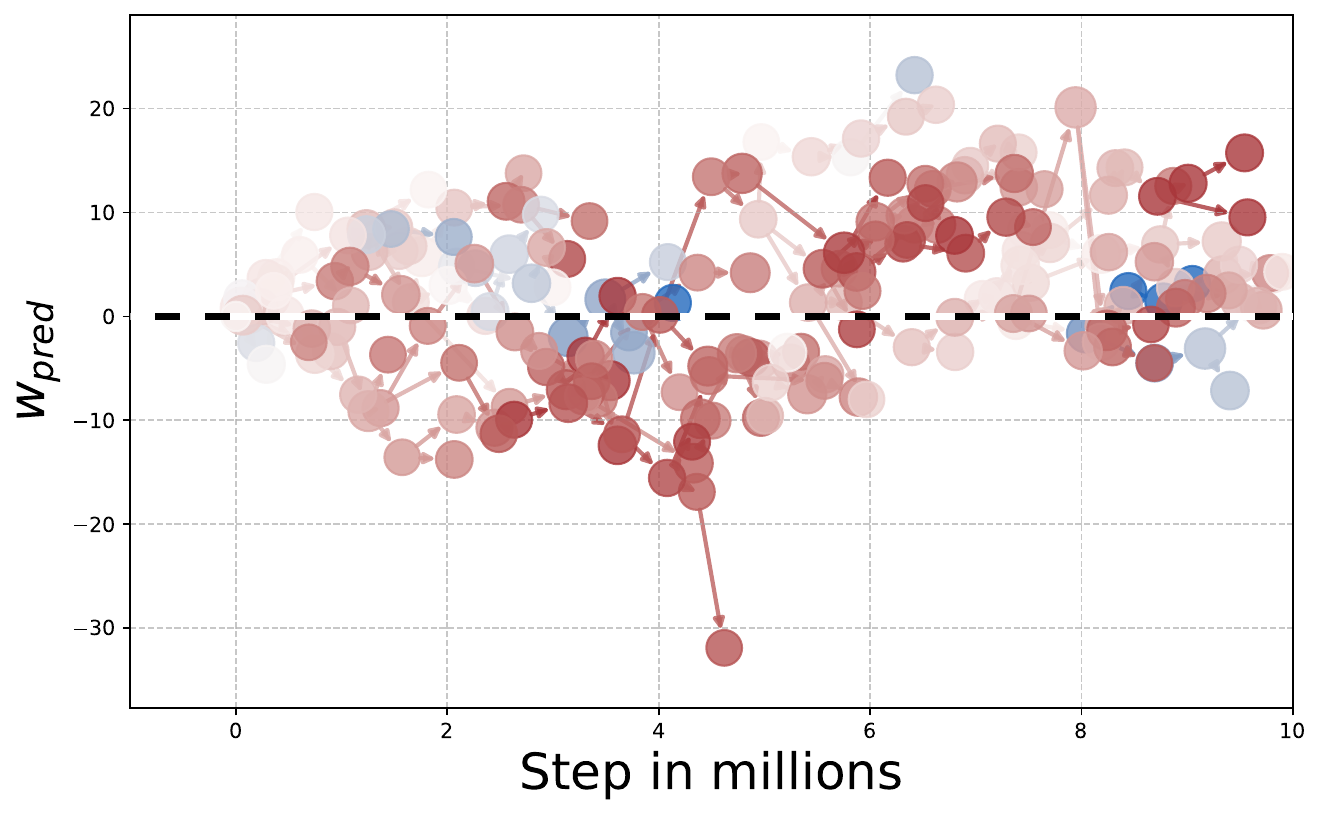}
    \caption{Seed 2}
    \label{fig:tree-2}
  \end{subfigure}
  \hfill
  \begin{subfigure}{0.23\linewidth}
    \centering
    \includegraphics[width=\linewidth]{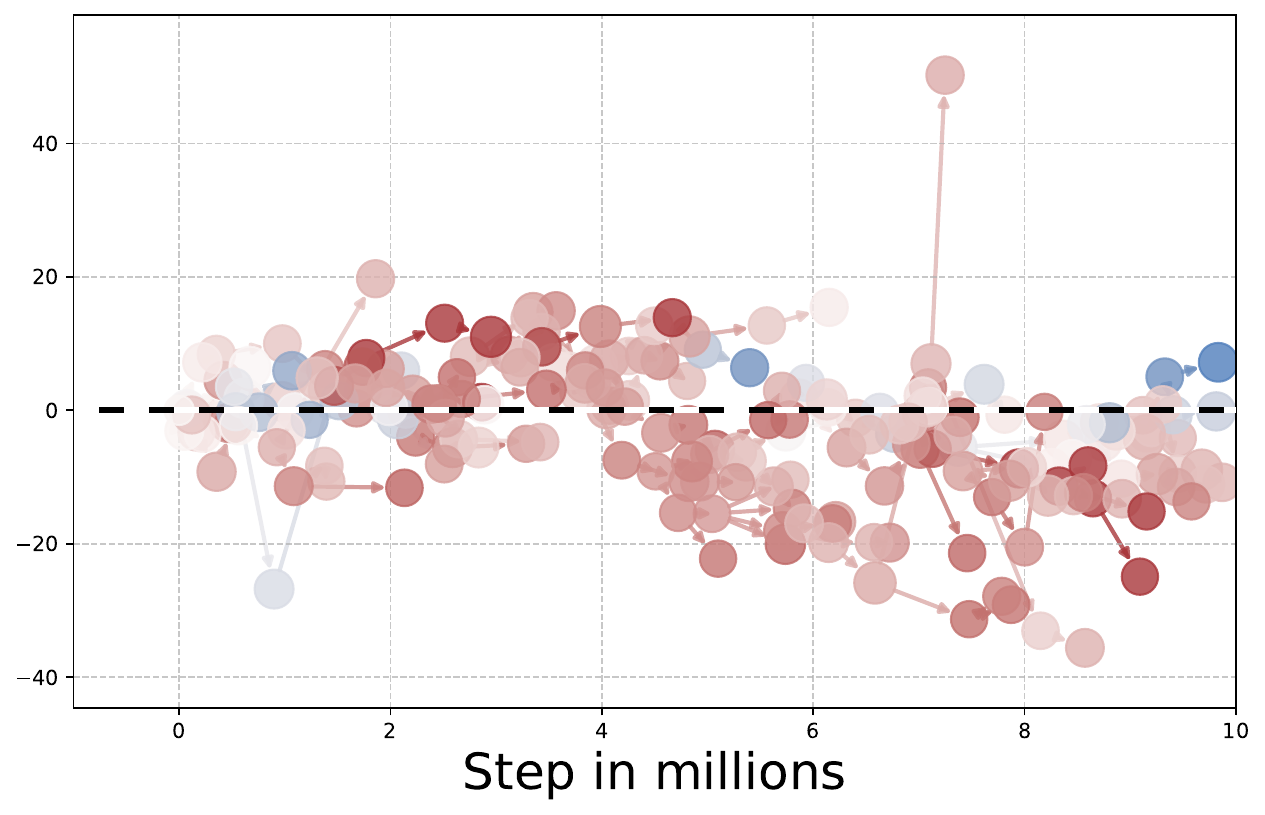}
    \caption{Seed 3}
    \label{fig:tree-3}
  \end{subfigure}
  \hfill
  \begin{subfigure}{0.23\linewidth}
    \centering
    \includegraphics[width=\linewidth]{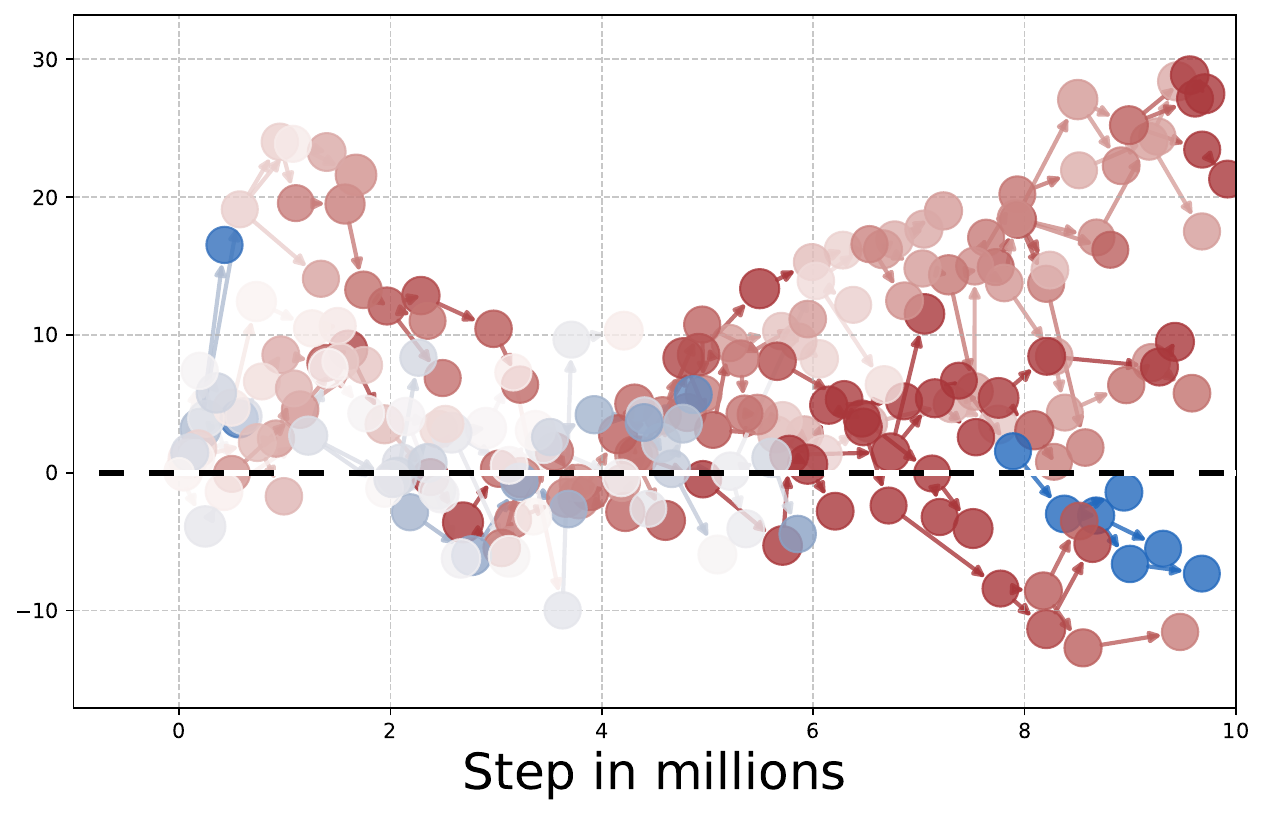}
    \caption{Seed 4}
    \label{fig:tree-4}
  \end{subfigure}
  \hfill
  \begin{subfigure}{0.27\linewidth}
    \centering
    \includegraphics[width=\linewidth]{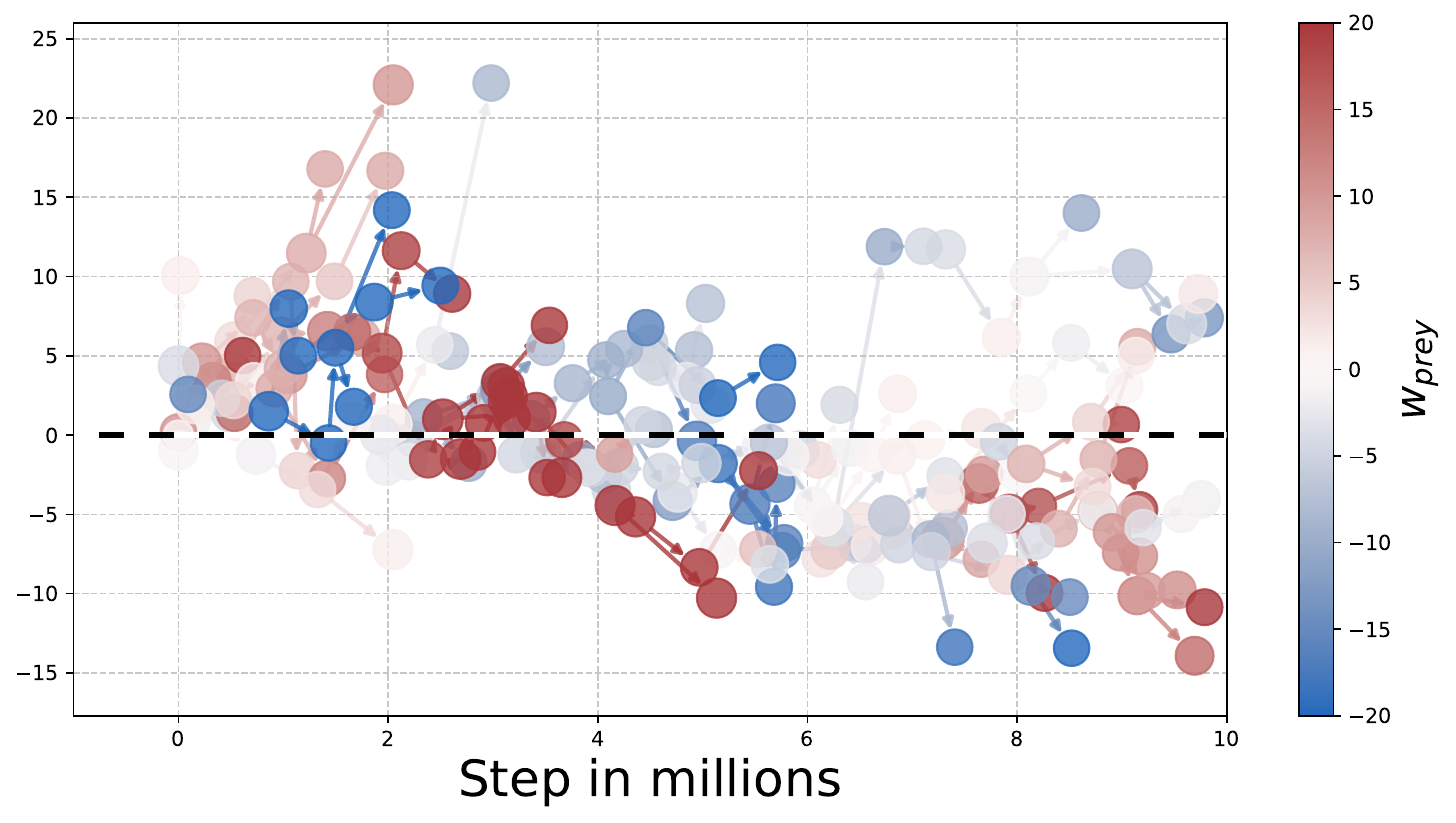}
    \caption{Seed 5}
    \label{fig:tree-5}
  \end{subfigure}
  \caption{Phylogenetic trees for prey across the remaining random seeds except seed 1, which is shown in \Cref{figure:tree}.}
  \label{figure:other-trees}
\end{figure*}

As illustrated in \Cref{figure:other-trees}, the phylogenetic trees for prey agents across the remaining random seeds show a consistent coexistence of individuals with both positive and negative $w_\mathrm{pred}$ values. While this diversity is maintained across all seeds, the dominant reward functions are different. Positive $w_\mathrm{pred}$ values are more frequent in seeds 2 and 4, whereas negative $w_\mathrm{pred}$ values are dominant in seeds 3 and 5.

\subsection{Dynamics Leading to Extinction}

\begin{figure}[t]
  \centering
  \includegraphics[width=7cm]{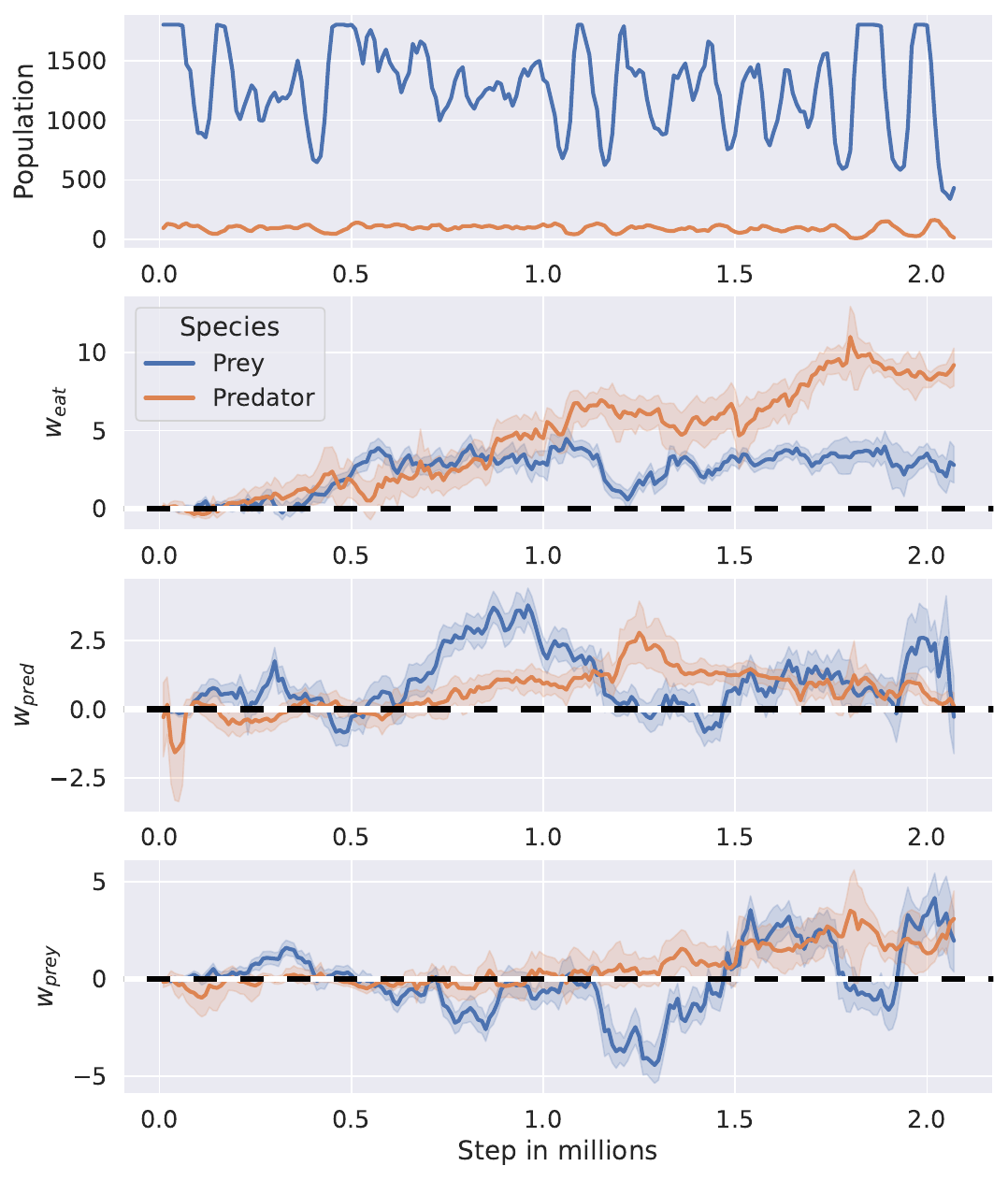}
  \caption{Population changes and the evolution of reward weights ($w_\mathrm{eat}$, $w_\mathrm{pred}$, and $w_\mathrm{prey}$) of prey and predators in the simulation run with large mouth and seed $4$, where predators went extinct after 2 million steps.}\label{figure:pop-extinction}
\end{figure}

We show the population dynamics and the evolution of $w_\mathrm{eat}$, $w_\mathrm{pred}$, and $w_\mathrm{prey}$ of prey and predators in the simulation run with large mouth and seed $4$ in \Cref{figure:pop-extinction}. We can see that predators went extinct after 2 million steps, following the sudden decrease of the prey population. We can see that the $w_\mathrm{eat}$ and $w_\mathrm{prey}$ for predators are high just before extinction, indicating that being too greedy is not a stable strategy for predators in this environment. At the same time, we can see the sudden drop of $w_\mathrm{pred}$ among prey agents, suggesting that many agents with larger $w_\mathrm{pred}$ got eaten around this period.

\end{document}